\documentclass[a4paper,fleqn]{cas-sc}

\usepackage[authoryear]{natbib}
\usepackage{lipsum}
\usepackage{longtable}
\newcommand{\flora}{FLoRA }

\usepackage{rotating}
\usepackage{array}
\usepackage{multirow}
\usepackage{booktabs}
\usepackage{makecell}

\usepackage{float} 
\usepackage{xcolor} 
\usepackage[utf8]{inputenc}

\usepackage{placeins}
\usepackage{flafter}
\def\tsc#1{\csdef{#1}{\textsc{\lowercase{#1}}\xspace}}
\tsc{WGM}
\tsc{QE}

\begin{document}
\let\WriteBookmarks\relax
\def\floatpagepagefraction{1}
\def\textpagefraction{.001}

\shorttitle{Facilitate Hybrid Human-AI Regulated Learning}    

\shortauthors{Xinyu Li}  

\title [mode = title]{FLoRA: An Advanced AI-Powered Engine to Facilitate Hybrid Human-AI Regulated Learning}  

\tnotemark[1]

\author[1]{Xinyu Li}[orcid=0000-0003-2681-4451]\ead{xinyu.li1@monash.edu}
\author[1]{Tongguang Li}[orcid=0000-0003-4616-5268]\ead{tongguang.Li@monash.edu}
\author[4,1]{Lixiang Yan}[orcid=0000-0003-3818-045X]\ead{lixiang.yan@monash.edu}
\author[5]{Yuheng Li}[orcid=0000-0002-5971-8469]\ead{yuheng.li@monash.edu}
\author[1]{Linxuan Zhao}[orcid=0000-0001-5564-0185]\ead{linxuan.zhao@monash.edu}
\author[1]{Mladen Rakovi\'c}[orcid=0000-0002-1413-1103]\ead{mladen.rakovic@monash.edu}
\author[3]{Inge Molenaar}[orcid=0000-0003-4639-2524]\ead{inge.molenaar@ru.nl}
\author[1]{Dragan Ga\v sevi\'c}[orcid=0000-0001-9265-1908]\ead{dragan.gasevic@monash.edu}
\author[2]{Yizhou Fan}[orcid=0000-0003-2777-1705]\ead{fyz@pku.edu.cn} \cormark[1]


\address[1]{Centre for Learning Analytics at Monash, Monash University, Melbourne, Australia}
\address[2]{Graduate School of Education, Peking University, Beijing, China}
\address[3]{Behavioural Science Institute, Radboud University, Nijmegen, Netherlands}
\address[4]{School of Education, Tsinghua University, Beijing, China}
\address[5]{Department of Applied Social Sciences, The Hong Kong Polytechnic University, Hong Kong, China}



\begin{abstract}
Self-Regulated Learning (SRL), defined as learners' ability to systematically plan, monitor, and regulate their learning activities, is crucial for sustained academic achievement and lifelong learning competencies. Emerging AI developments profoundly influence SRL interactions by potentially either diminishing or strengthening learners' opportunities to exercise their own regulatory skills. Recent literature emphasizes a balanced approach termed Hybrid Human-AI Regulated Learning (HHAIRL), in which AI provides targeted, timely scaffolding while preserving the learners' role as active decision-makers and reflective monitors of their learning process. Central to HHAIRL is the integration of adaptive and personalized learning systems; by modelling each learner’s knowledge and self-regulation patterns, AI can deliver contextually relevant scaffolds that support learners during all phases of the SRL process. Nevertheless, existing digital tools frequently fall short, lacking adaptability and personalisation, focusing narrowly on isolated SRL phases, and insufficiently supporting meaningful human-AI interactions. In response, this paper introduces the enhanced \flora Engine, which incorporates advanced generative AI features and state-of-the-art learning analytics, and grounds in solid educational theories. The \flora Engine offers tools such as collaborative writing, multi-agent chatbots, and detailed learning trace logging to support dynamic, adaptive scaffolding of self-regulation tailored to individual needs in real time. We further present a summary of several research studies that provide the validations for and illustrate how these tools can be utilized in real-world educational and experimental contexts. These studies demonstrate the effectiveness of \flora Engine in fostering SRL, providing both theoretical insights and practical solutions for the future of AI-enhanced learning contexts.

\end{abstract}


\begin{highlights}
\item The \flora Engine provides advanced instrumentation tools to effectively detect, measure, and facilitate self-regulated learning (SRL) processes.
\item Leveraging GenAI and learning analytics, \flora Engine delivers personalized scaffolds that dynamically support diverse learners throughout the SRL cycle.
\item The \flora Engine exemplifies the potential of AI-enhanced platforms to address the complexities of hybrid human–AI regulation in SRL, as demonstrated across multiple educational case studies.
\end{highlights}


\maketitle

\section{Introduction}\label{sec1} 


Self-regulated learning (SRL) is the ability to plan, monitor, and adjust one’s learning by coordinating cognitive, metacognitive, motivational, and emotional processes~\citep{azevedo2023theories,du2023can}. Learners with strong SRL set clear goals, track their progress, and adapt strategies, which typically leads to better outcomes across tasks and over time~\citep{xu2023meta,zhao2025meta}. Given the importance of SRL and the rapid evolution of digital learning environments and tools, it is crucial to reconsider how SRL is understood and supported~\citep{seufert2018interplay,winne2017learning}. In this context, advances in Generative Artificial Intelligence (GenAI) are particularly significant, as they are beginning to reshape how educators, researchers, and learners conceptualize and foster SRL~\citep{jarvela2023human}. On the one hand, many adaptive learning technologies can offload SRL processes like monitoring and selecting new materials, which can inadvertently reduce learners’ opportunities to exercise self-regulation~\citep{MOLENAAR2023107540}. On the other hand, the integration of data-driven AI support with explicit scaffolds aimed at promoting and reinforcing learners' own regulatory skills~\citep{molenaar2022concept,lim2024students}. The AI can step in with timely, data-driven insights while still encouraging learners to make key decisions and reflect on their learning process~\citep{molenaar2022temporal}. This synergy leverages the strengths of both AI and the human learner, ensuring that the learner remains at the centre and continues to build the metacognitive monitoring and control skills essential for future learning, with the process termed as Hybrid Human-AI Regulated Learning (HHAIRL)~\citep{huang2024promoting, nguyen2025human, jarvela2023human,molenaar2022temporal}. Still, there is a risk that an excessive reliance on AI could diminish learners’ engagement in important self-regulatory activities \citep{zhang2024systematic}, such as monitoring their own progress or planning their learning, thereby impeding the growth of robust SRL skills~\citep{fan2025beware}.

Despite increasing attention to Hybrid Human-AI Regulation~\citep{huang2024promoting, nguyen2025human, jarvela2023human,molenaar2022temporal}, several gaps and challenges remain. First, existing technologies for the real-time identification of a learner’s SRL are relatively underdeveloped. Many current tools lack adaptability; they often work well for specific tasks but do not account for the diverse realities of educational environments, where students engage with varied materials and may frequently learn collaboratively~\citep{alvarez2022tools}. In addition, many tools focus on particular SRL phases rather than encompassing the entire cycle of setting goals, monitoring progress, and evaluating outcomes~\citep{alvarez2022tools,guan2025educational}. The design of tools like dashboards that help learners understand and exert control alongside the AI is still at an early stage; simply showing performance metrics is not enough~\citep{matcha2019systematic}. 
Furthermore, the emergence of AI especially GenAI, capable of providing human-like interactions and demonstrations of expert-level reasoning, significantly alters typical human-tool interaction patterns~\citep{edwards2025human, kasneci2023chatgpt}. Learners increasingly trust and rely on GenAI systems due to their sophisticated capacities and human-like interactions, fundamentally changing how learners engage in SRL processes such as goal-setting, monitoring, and reflection \citep{hauske2024can,lai2024adapting}. Nonetheless, existing SRL support tools, primarily designed prior to these advancements, are insufficiently equipped with GenAI-based functionalities to reliably track, interpret, and scaffold these complex human-GenAI interactions~\citep{edwards2025human, holmes2023evaluating, tsai2025effects}. Finally, many SRL support tools often exhibit strong dependencies on their specific learning contexts~\citep{wong2019supporting, matcha2019systematic, geng2025effects}, limiting their applicability across diverse disciplines and making them unsuitable for evaluating the effectiveness of HHAIRL in diversified learning scenarios~\citep{molenaar2022concept}.

In response to these limitations, the \flora Engine has been developed based on the previous work~\citep{li2025floraengine}. Building upon prior version, the current version of \flora not only significantly expands the three main modules -- instrumentation tools, trace parser, and scaffolding, but also introduces several innovative features to support HHAIRL research. One particularly significant advancement in this iteration is the integration of GenAI technologies, such as GPT-4o, which enable more dynamic, personalised, and adaptive learner-centred support. In parallel, \flora incorporates advanced learning analytics methodologies~\citep{gavsevic2015let,matcha2019detection} grounded firmly in established SRL theories~\citep{saint2020trace,siadaty2016trace}, ensuring that insights generated from learner-interaction data remain theoretically coherent and practically meaningful. By aligning closely with emerging trends in AI-enhanced educational practice, \flora serves a dual purpose: on one hand, it provides researchers with a comprehensive, robust platform for systematically investigating the complex dynamics of Hybrid Human-AI regulation; on the other, it offers educators and learning scientists a practical example illustrating how GenAI technologies can effectively enhance SRL-supportive educational contexts.

In the following sections, a detailed description of \flora Engine's features and design rationale are presented. Subsequently, empirical evidence gleaned from illustrative case studies are described, clearly demonstrating how these novel features concretely support learners' SRL capabilities, across diverse, AI-mediated educational scenarios.

\section{Theoretical Foundations}

\subsection{Self-Regulated Learning}
Self-regulated learning (SRL) is a comprehensive framework that encompasses the cognitive, metacognitive, behavioural, motivational, and emotional aspects of learning~\citep{zimmerman1986becoming,winne2018theorizing,winne1998studying,winne2017cognition}. Through a range of strategies, such as planning, goal setting, self-monitoring, self-instruction, and self-evaluation, SRL enable learners to take control of their own learning processes. By engaging in SRL, learners become proactive in their education, employing various tactics to understand and master new information, while also managing their emotions and staying motivated throughout the learning journey~\citep{panadero2017review}.

The importance of SRL lies in its significant impact on educational outcomes and lifelong learning~\citep{dent2016relation,xu2023meta,broadbent2015self,guo2022using}. SRL equips learners with the skills necessary to adapt to various learning contexts and challenges, fostering autonomy and resilience. When students practice SRL, they are more likely to exhibit higher levels of self-efficacy, persistence, and academic achievement~\citep{uzir2020analytics,van2022dynamics}. Moreover, SRL promotes deeper engagement with learning materials~\citep{li2022cognitive}, critical thinking~\citep{anwar2022exploration}, and the ability to transfer knowledge to new situations~\citep{stebner2022transfer,schuster2020transfer}. As educational environments become increasingly complex, the ability to self-regulate is essential for learners to succeed in learning performance. Educators recognise that teaching students how to learn, rather than just what to learn, empowers them to become effective, lifelong learners who can navigate and thrive in a rapidly changing world~\citep{guo2022using}.


\subsection{The Detection and Measurement of SRL}

Detecting and measuring SRL is pivotal for understanding how learners control and direct their learning processes, especially in computer-based environments~\citep{molenaar2023measuring}. SRL involves a spectrum of cognitive, metacognitive, motivational and affective processes~\citep{azevedo2015defining,winne2023roles}, such as setting goals, monitoring progress, and adjusting strategies to achieve learning objectives. However, capturing these internal processes poses significant challenges because they are often not directly observable and may not always manifest in overt behaviours that can be easily recorded or quantified~\citep{saint2020combining}.

One effective approach to detecting SRL is the utilization of log data and learning analytics within digital learning platforms~\citep{siadaty2016trace,winne2022modeling,bernacki2025leveraging}. As learners interact with computer-based educational resources, they generate extensive trace data—clickstreams, navigation paths, and tool usage logs—that can be analysed to infer SRL activities~\citep{papamitsiou2014learning, bannert2014process,azevedo2019analyzing,li2020using,deininger2025using}. For instance, if a learner chooses to attempt a quiz before engaging with reading materials, this might indicate proactive planning and goal-setting behaviours. However, not all SRL actions leave discernible traces in the log data, and some SRL processes, particularly metacognitive ones, may remain hidden or indistinct, making it challenging to achieve a comprehensive understanding solely through raw interaction data~\citep{winne2017cognition,winne2018theorizing,geng2025effects}.

To address limitations in trace-based measurements of SRL, researchers have developed and integrated instrumentation tools within learning environments that are specifically designed to elicit and capture SRL processes~\citep{winne2019nstudy,van2021instrumentation}. These tools serve a dual purpose: they facilitate learning by providing functionalities that support SRL, and they generate detailed trace data that reflect learners' cognitive and metacognitive activities. For example, tools that allow learners to highlight text, take notes, or set reminders encourage them to engage in organization, elaboration, and planning—key components of SRL~\citep{paans2019temporal, cerezo2020process}. A pioneering exemplar of such a learning technology is nStudy~\citep{winne2019nstudy}, which offers a suite of features enabling learners to interact deeply with content while simultaneously logging these interactions for analysis. From the learners’ perspective, these tools empower them to actively engage in self-regulation by making their cognitive strategies explicit. For educators and researchers, the instrumentation tools provide valuable data that can be analysed to gain insights into the learners' SRL processes.

Analysing the rich trace data generated requires sophisticated methods to accurately map observable behaviours to underlying SRL processes~\citep{bannert2007metakognition,molenaar2023measuring}. Conceptualizing SRL as a series of events with increasing complexity—occurrence, contingency, and patterned contingency—provides a structured framework for this analysis~\citep{lim2021temporal}. The occurrence level captures the frequency of specific learning actions, such as the number of times a learner highlights text. The contingency level examines sequences of actions, providing context by looking at how one action follows another—for example, highlighting text immediately after taking a quiz. The patterned contingency level involves identifying recurring sequences or patterns of actions that suggest a strategic approach to learning, such as a consistent cycle of reading, note-taking, and self-testing. By employing temporal analytical methods like process mining and Markov models, researchers can uncover these patterns within the data, revealing how learners self-regulate over time~\citep{saint2022temporally,li2025analytics}.

\subsection{Hybrid Human-AI and Socially Shared Regulated Learning} 

Socially-Shared Regulation of Learning (SSRL) is the process through which members of a group collectively plan, monitor and adapt their activity in order to reach a shared learning goal~\citep{jarvela2018contemporary,haataja2022individuals}. Regulation decisions—clarifying the task, setting sub-goals, checking mutual understanding or reallocating effort—are negotiated and enacted together, not left to any single individual~\citep{zheng2019effects,kent2020investigating}. Traditionally, teachers have scaffolded these collective metacognitive moves; now GenAI can join the conversation~\citep{jarvela2023human,yan2025effects,tsai2025effects}. By analysing multimodal data streams in real time, an AI agent can detect moments when the group drifts apart, highlight emerging misconceptions or surface silent contributions, effectively acting as an additional “group member” that nudges the team toward productive dialogue~\citep{edwards2025human}. Thus, AI becomes an enabler of richer SSRL, augmenting human awareness rather than replacing it.

Hybrid Human-AI Regulation (HHAIR) generalises this idea by treating regulation itself as a shared responsibility that can fluidly shift between learners and intelligent systems~\citep{molenaar2022concept}. A hybrid system begins by off-loading some regulatory work to the AI—diagnosing knowledge gaps, recommending resources, or visualising progress—then gradually on-loads control back to the learner or the group as their self-regulatory competence grows~\citep{nazaretsky2024ai, zhang2024systematic}. Advances in GenAI amplify this potential: large language models (LLMs) can provide context-sensitive explanations, create adaptive prompts, or summarise a group’s discussion on demand~\citep{whalen2023chatgpt}. At the same time, AI can automate repetitive assessment tasks and mine vast data sets for patterns invisible to teachers, freeing educators to devote their time to higher-order coaching and socio-emotional support~\citep{cukurova2024interplay}. The essence of HHAIR is this dynamic handover of control, calibrated to learners’ needs and aimed at cultivating durable self- and co-regulation skills, which aims to the final goal--AI Regulation--as explained in Table~\ref{tab:hybrid}. HHAIR emphasises the importance of human agency and oversight even when using AI assistance~\citep{lai2022trends}.


\begin{table}[ht]
\centering
\renewcommand{\arraystretch}{1.5} 
\caption{Hybrid Human-AI Regulation Levels~\citep{molenaar2022concept}}\label{tab:hybrid}
\begin{tabular}{|>{\hspace{0pt}}m{0.145\linewidth}|>{\hspace{0pt}}m{0.3\linewidth}|>{\hspace{0pt}}m{0.42\linewidth}|} 
\hline
Degrees of hybrid regulation & Description & Instrumentation Tool Examples \\ 
\hline
AI regulation & AI monitors and adjusts extensively, with humans aware of its regulation & Tools that are fully controlled by AI, which adapts learning processes and content while keeping learners informed of its regulatory actions to support their learning. \\ 
\hline
Co-regulation & AI monitors and adjusts incrementally, and humans understand its monitoring and control & Tools that collect multimodal learner data, such as facial expressions, EDA, eye-tracking, and learning traces, and use this data to adapt learning content and provide scaffolding in real-time \\ 
\hline
Shared-regulation & AI monitors and suggests control actions, while humans understand monitoring and perform control & Tools that access learners’ written work, use AI for analysis, and deliver feedback~\citep{tang2024facilitating} \\ 
\hline
Self-regulation & Humans monitor and initiate control independently & Tools for creating plan and checking time~\citep{winne2019nstudy}  \\
\hline

\end{tabular}

\end{table}

Attending to SSRL and HHAIR is therefore crucial in contemporary education. Research shows that learners often under-utilise effective regulation strategies, while many adaptive tools silently take those decisions away, depriving students of practice~\citep{nazaretsky2024ai}. Thoughtfully designed AI tools and systems need to address both problems: it keeps collaboration efficient in the moment, provides personalised and timely feedback, and explicitly models the metacognitive moves for learners to internalise~\citep{cukurova2024interplay,yan2024promises}. However, these tools remains limited. One key reason for this gap is the complexity of designing adaptive AI systems that can dynamically share control with human learners and educators~\citep{hao2023exploring}. Current AI-driven educational technologies are heavily focused on providing automated and fixed recommendations, often lacking mechanisms for reciprocal interaction and real-time feedback integration from learners~\citep{roll2015understanding}. Most existing adaptive learning systems prioritise knowledge acquisition and task adaptation, rather than supporting SRL and metacognitive processes, which require the AI to recognise, interpret, and respond to a wide range of learner behaviours, emotions, and cognitive states dynamically~\citep{azevedo2019analyzing}. Developing systems that can contextually adjust the level of intervention based on individual's skills and autonomy remains a major technical challenge. This intervention means the support offered by the systems such as feedback of learners' current learning progress or scaffolding that can lead to cognitive and metacognitive learning process adjustments~\citep{saint2024analytics}. Furthermore, most learning management systems (LMS) and AI-driven tutoring platforms lack transparent, explainable AI mechanisms to ensure that learners and teachers can understand how AI-driven decisions are made~\citep{shin2021effects,khosravi2022explainable}. Without clear communication of AI reasoning, users may struggle to trust or effectively engage with the system~\citep{kizilcec2016much}. Limited interdisciplinary collaboration among AI developers, cognitive scientists, and education practitioners further contributes to the slow development of robust hybrid regulation tools~\citep{molenaar2021personalisation}. 

While AI holds immense promise in education—such as supporting SSRL and HHAIR with personalised feedback and guidance—recent studies caution that heavy reliance on AI may lead to metacognitive disengagement, where learners offload reflective thinking to the AI. For example, \cite{zou2024investigating} found that students were more engaged and made more accurate revisions when using human teacher feedback compared to feedback from ChatGPT, suggesting that AI assistance, while helpful, might not stimulate the same level of active effort. Potential risks of intensive AI use in learning include a loss of student agency where learners may become passive and allow AI to make decisions for them~\citep{lepage2024preserving}; the excessive automation of metacognitive processes which can prevent learners from practising self-monitoring or critical thinking because AI takes over these functions~\citep{gerlich2025ai,zhai2024effects}; and the reproduction of biases, as AI models may mirror or even amplify existing societal biases in their feedback or suggestions, potentially leading to unfair or skewed learning experiences~\citep{chinta2024fairaied,suchithra2025study}.



The \flora Engine has been developed and will continue to evolve to address these challenges by harnessing expertise from multiple areas, including researchers, educators, and AI developers. Its primary objective is to bridge the gap between AI and SRL, fostering a more adaptive and personalised learning experience. To achieve this, \flora provides a diverse set of tools designed to support both teachers and students. Teachers can utilise these tools to design and customise course content, while also maintaining control over the degree of AI intervention in the learning process. Meanwhile, students receive personalised support to enhance their SRL skills. Additionally, learner and teacher feedbacks are continuously integrated into the system, ensuring ongoing improvements based on real-world educational interactions. Ultimately, \flora aspires to develop fully autonomous AI Regulation (explained in Table~\ref{tab:hybrid}), wherein the system dynamically adjusts its support based on each learner’s progress, thereby maximizing the effectiveness of AI in education.


\section{Innovative Features}

This section introduces the major advancements incorporated into the \flora Engine, emphasizing the core innovations that distinguish \flora from existing tools and approaches. Particular attention is given to the most notable instrumentation tools developed, with detailed accounts of their functionalities and underlying process logic. These enhancements demonstrate the capacity of \flora to serve as a state-of-the-art research tool for investigating and supporting SRL within rapidly evolving educational environments shaped by the widespread adoption of GenAI technologies.

\begin{figure}[ht]
\centering
\includegraphics[width=0.85\linewidth]{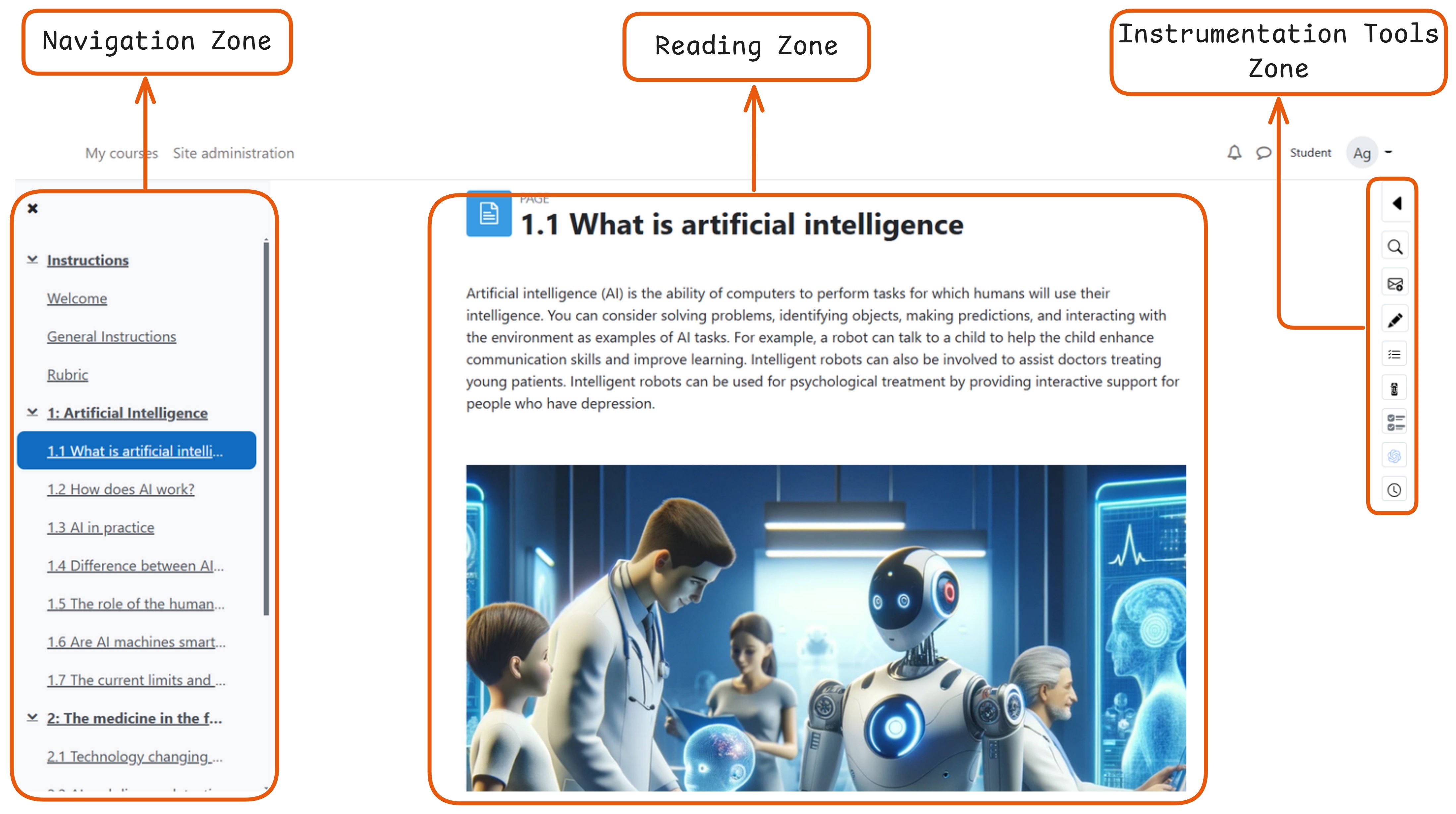}
\caption{Overview of the \flora User Interface}
\label{fig:flora_interface_overview}
\end{figure}

Figure~\ref{fig:flora_interface_overview} presents the integration of the \flora Engine within the Moodle platform, highlighting its user interface. Three principal zones are identified: the Navigation zone, Reading zone, and Instrumentation tools zone, with the latter being offered by the \flora Engine. While the Navigation and Reading zones are standard Moodle components, the Instrumentation tools zone enables comprehensive tracking and processing of all learner interactions within the web pages.

\subsection{From Learning Analytics to Personalised Scaffolds with GenAI}
\label{subsec:GenAI-scaffolds}
\begin{figure}[ht]
\centering
\includegraphics[width=0.85\linewidth]{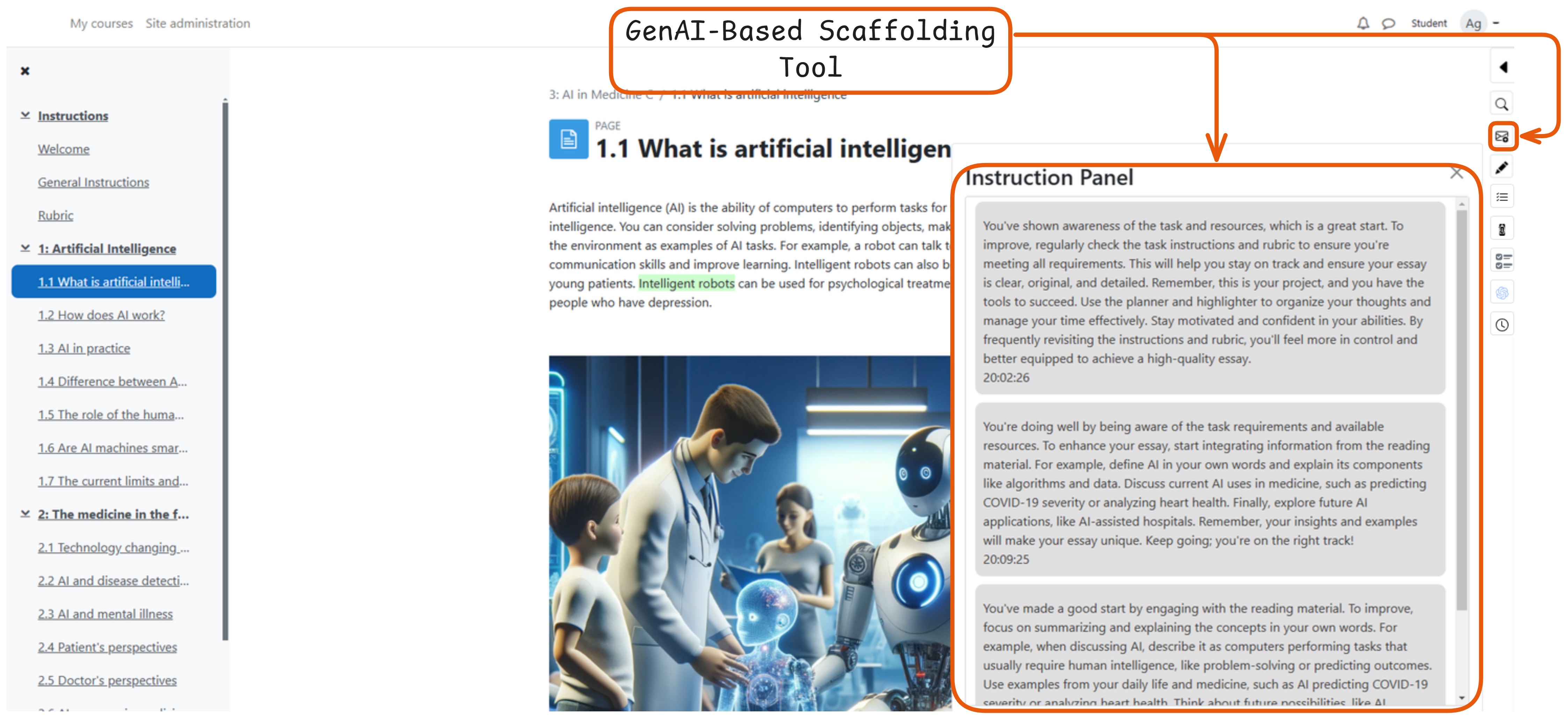}
\caption{GenAI-based scaffolding tool}
\label{fig:gpt_scaffolding_tool}
\end{figure}

\flora Engine incorporates an analytics-based GenAI scaffolding function that integrates learning analytics and GenAI to offer personalised support for students’ SRL processes. The tool is shown in Figure~\ref{fig:gpt_scaffolding_tool}. Its name has been updated to "Instruction Panel" to enhance user clarity. At predefined intervals, users receive scaffolding messages through this panel, providing personalised guidance during their learning process. 

\begin{figure}[ht]
\centering
\includegraphics[width=0.85\linewidth]{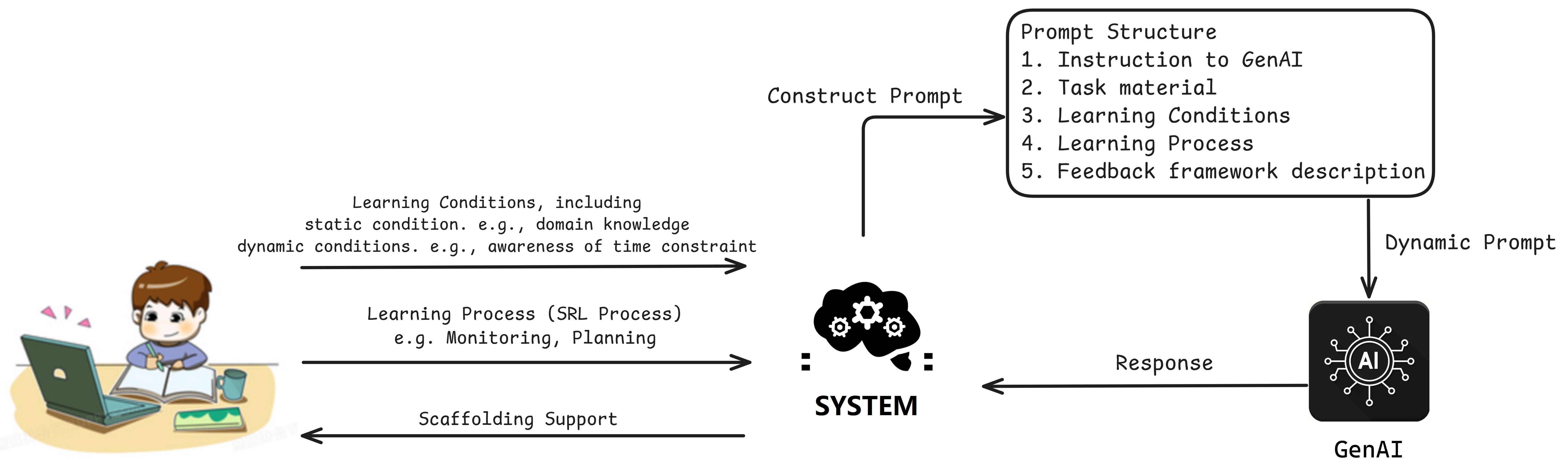}
\caption{GenAI-based scaffolding trigger mechanism}
\label{fig:gpt_scaffolding_trigger}
\end{figure}

The mechanism that triggers this support is illustrated in Figure~\ref{fig:gpt_scaffolding_trigger}. As learners work on the given tasks, learning analytics techniques collect real-time data, capturing both internal learning conditions and SRL processes. According to \citet{winne2022modeling}, internal conditions comprise, for instance, a learner’s level of domain knowledge (a static condition) and their awareness of time constraints (a dynamic condition). In a time-limited learning task, a learner might initially adopt a linear approach, but as they become aware of the remaining time, they may adjust their strategies to accommodate this constraint. Consequently, time-awareness can be considered a dynamic learning condition that updates over the course of the task. In \flora, such changes in dynamic conditions are detected in real-time through trace data (e.g., a click on the timer to check the remaining time). Meanwhile, static conditions, such as learners’ domain knowledge levels, are determined through pre-task surveys and assessments. \flora automatically scores these instruments, and the resulting scores feed into the GenAI-based scaffolding tool to tailor the support provided to each learner. Each learning condition (defined in Table~\ref{tab:conditions}) is converted into a predefined statement and added to the GenAI prompt. More details can be found in Appendix~\ref{appendix_GPT_scaffold_trigger_logic}. For example, if a student's trace data shows they clicked on the timer, it indicates awareness of time constraints, generating “the student is aware of the time constraint.” Conversely, no click indicates unawareness, resulting in “the student is not aware of the time constraint.”

SRL processes are continuously recorded in real time, providing insights into how learners actively manage and regulate their learning activities. The current version of \flora is designed to be compatible with various theoretical frameworks that inform the development of the trace parser. Notably, this includes Winne and Hadwin’s COPES model \citeyearpar{winne2022modeling}, as well as Bannert’s framework \citeyearpar{bannert2007metakognition}, which builds upon Zimmerman’s cyclical model of SRL \citeyearpar{zimmerman2013cognitive}. To illustrate, consider an example based on Bannert’s framework: when a student is engaged in drafting an essay, and subsequently reviews the task instructions and rubrics before returning to continue writing, the raw interaction trace records these learning activities as \textit{WRITE\_ESSAY} and \textit{READ\_INSTRUCTION}. The sequential pattern \textit{WRITE\_ESSAY → READ\_INSTRUCTION → WRITE\_ESSAY} is then annotated as representing the Metacognition-Evaluation phase of the SRL process. A comprehensive explanation of how raw trace data are mapped onto specific SRL processes is provided by \cite{li2025floraengine}, and this mapping procedure has been empirically validated using think-aloud protocols~\citep{fan2022towards}. Further, a recent laboratory study demonstrated that certain SRL processes are significantly correlated with improved learning outcomes at specific stages of task completion~\citep{li2024analytics}. For example, in a timed reflective writing task, students who engaged in orientation behaviours—such as consulting task instructions and assessment rubrics within the first two minutes—achieved significantly higher scores than those who did not~\citep{van2023design}. Building upon these findings, \flora has been designed to automatically trigger GenAI-powered scaffolding whenever critical SRL processes are not detected within a predefined time window, thereby providing timely and targeted support to enhance learning. The thresholds that trigger scaffolding are not fixed but are configurable and defined according to the specific design and aims of each study.

\begin{figure}[ht]
\centering
\includegraphics[width=0.65\linewidth]{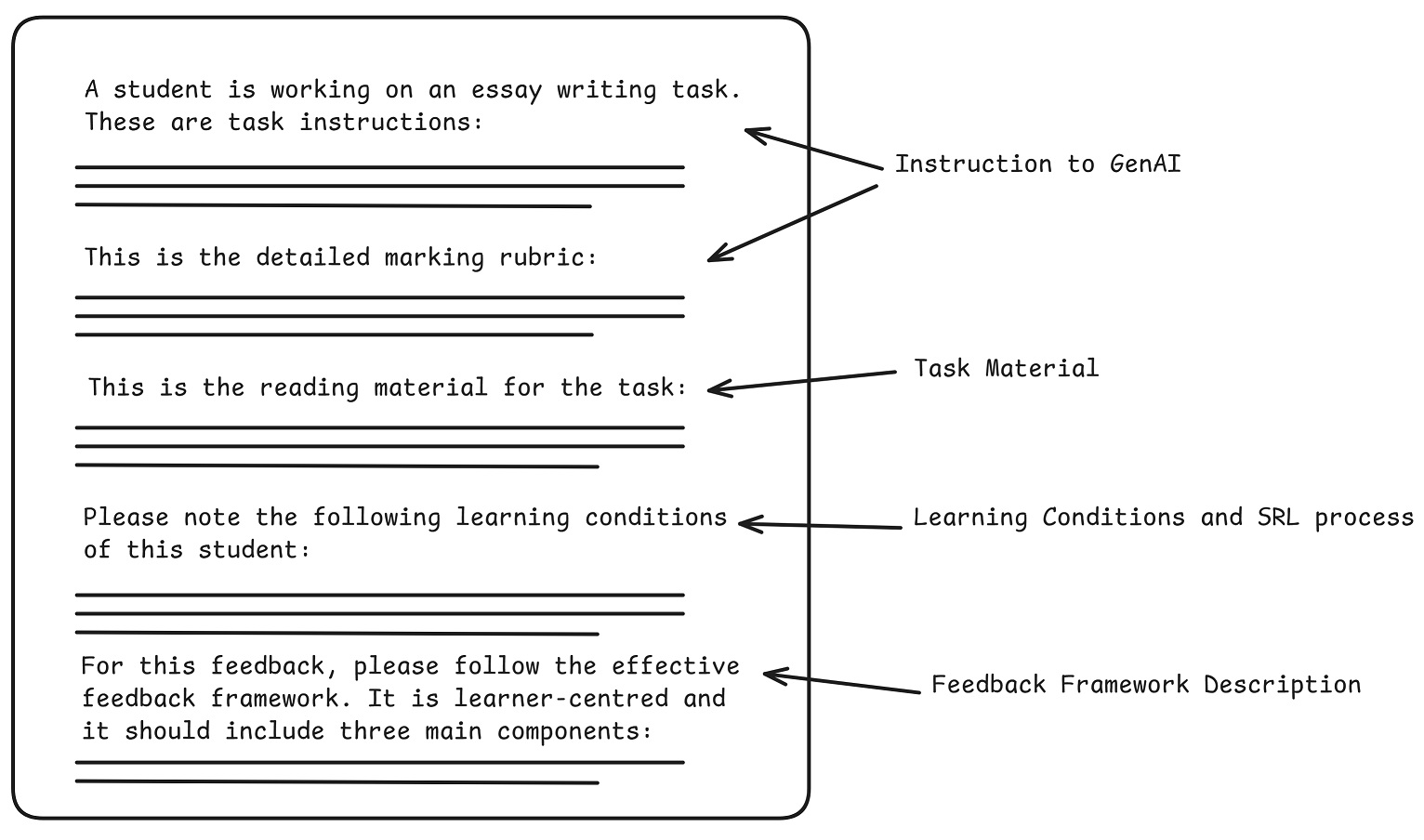}
\caption{GenAI Prompt Example}
\label{fig:genai_prompt_example}
\end{figure}

As shown in Figure \ref{fig:gpt_scaffolding_trigger}, learning conditions and critical SRL processes—collected through real-time analytics—are sent to the \flora system, which automatically generates a GenAI prompt using the relevant template. An example of a GenAI prompt is presented in Figure \ref{fig:genai_prompt_example}. In this template, the "Learning Conditions and SRL Process" section is dynamically generated, while the remaining components remain consistent across all instances. These prompts are then passed to GenAI to produce responses that become SRL scaffolding. The resulting feedback is delivered to students via the GenAI-based scaffolding tool, nudging them toward effective SRL processes with content tailored to their evolving needs (Figure \ref{fig:gpt_scaffolding_tool}). This integrated design—combining learning analytics with GenAI—offers clear advantages over rule-based scaffolding, which relies on a predetermined set of messages \citep{srivastava2022effects, van2023design}. Since rule-based systems require all potential messages to be predefined, addressing multiple learning conditions simultaneously becomes unwieldy and impractical due to the sheer complexity of enumerating possible rule combinations \citep{backhaus2017assessing}.



\begin{figure}[ht]
\centering
\includegraphics[width=0.85\linewidth]{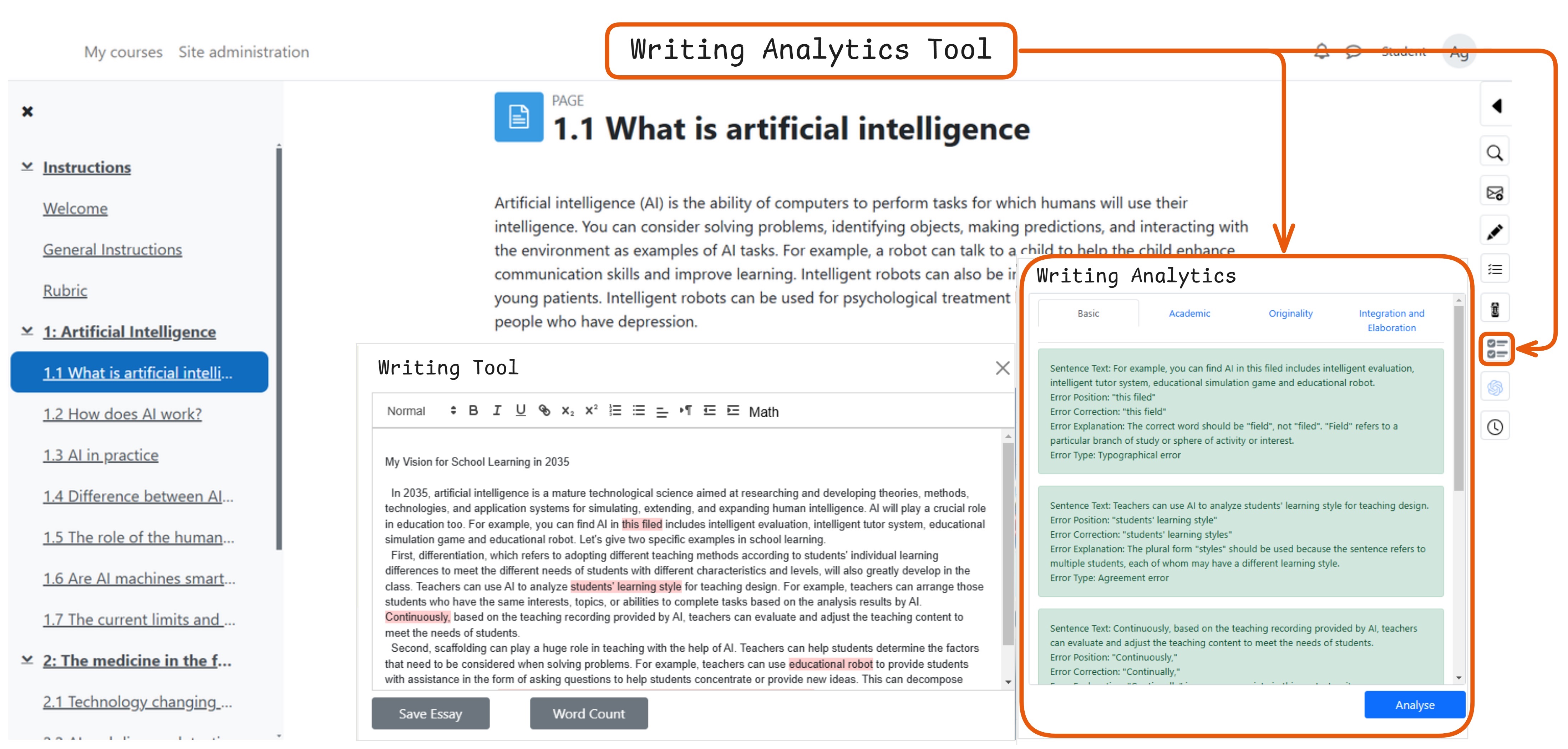}
\caption{Writing analytics tool - basic writing}
\label{fig:writing_analytics_tool_grammar}
\end{figure}

\subsection{From Writing Analytics to Personalised Feedback}\label{subsec:writing-analytics}

The writing analytics tool in \flora is specifically designed to foster SRL within the context of information problem-solving tasks. By delivering targeted, formative feedback for revision, the tool cultivates essential SRL processes such as self-monitoring, reflection, and strategic adaptation. Its four core functions address different dimensions of writing proficiency, enabling iterative improvement and the development of critical metacognitive skills.

\begin{figure}[ht]
\centering
\includegraphics[width=0.85\linewidth]{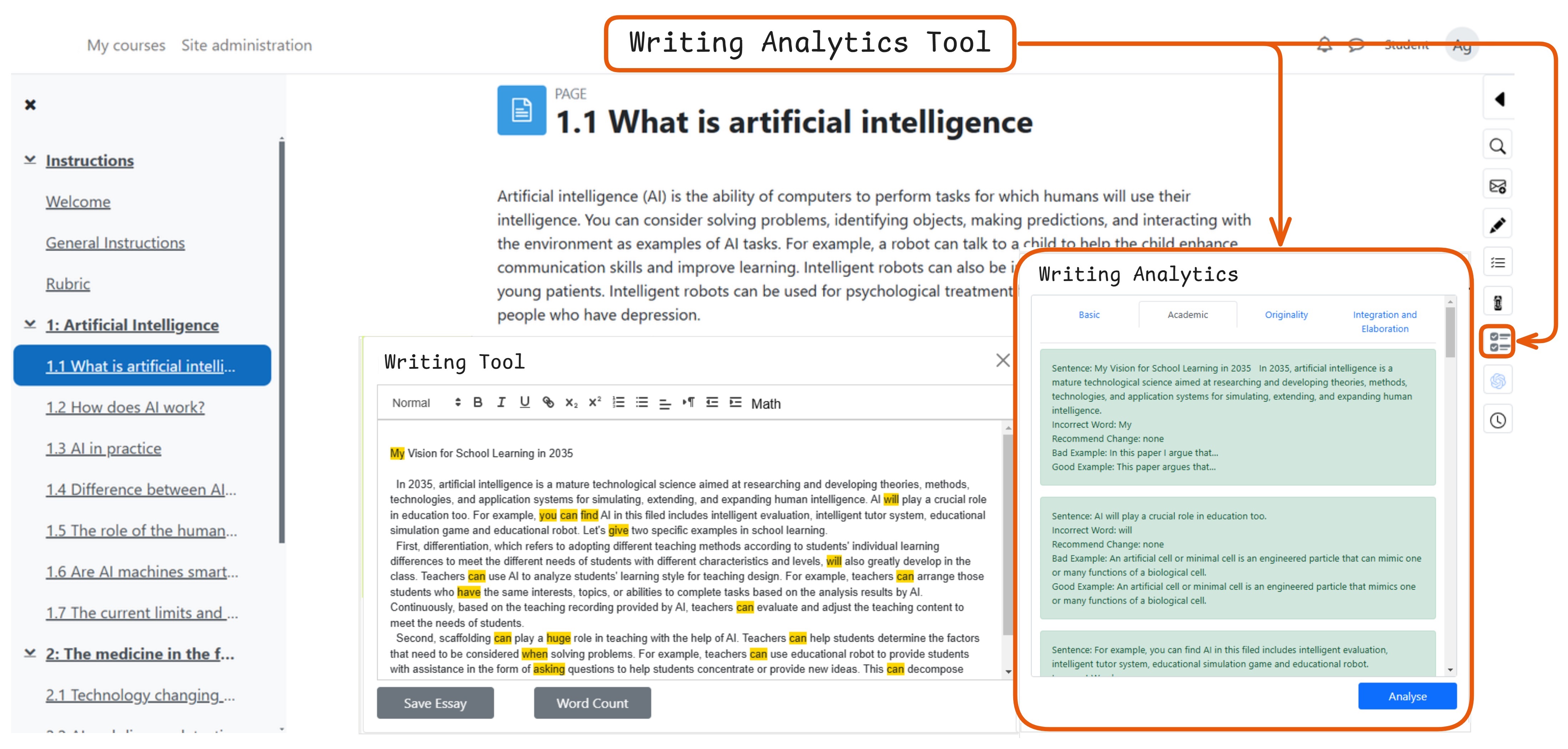}
\caption{Writing analytics tool - academic writing}
\label{fig:writing_analytics_tool_vocab}
\end{figure}

The first core function, the Basic Writing module (see Figure~\ref{fig:writing_analytics_tool_grammar}), focuses on identifying and correcting spelling and grammatical errors. Feedback is generated through GenAI assessments, ensuring accuracy and specificity in suggestions. By addressing foundational language issues, this module not only enhances grammatical correctness but encourages students to regularly evaluate and refine their writing—a key aspect of the SRL process of monitoring.

\begin{figure}[ht]
\centering
\includegraphics[width=0.85\linewidth]{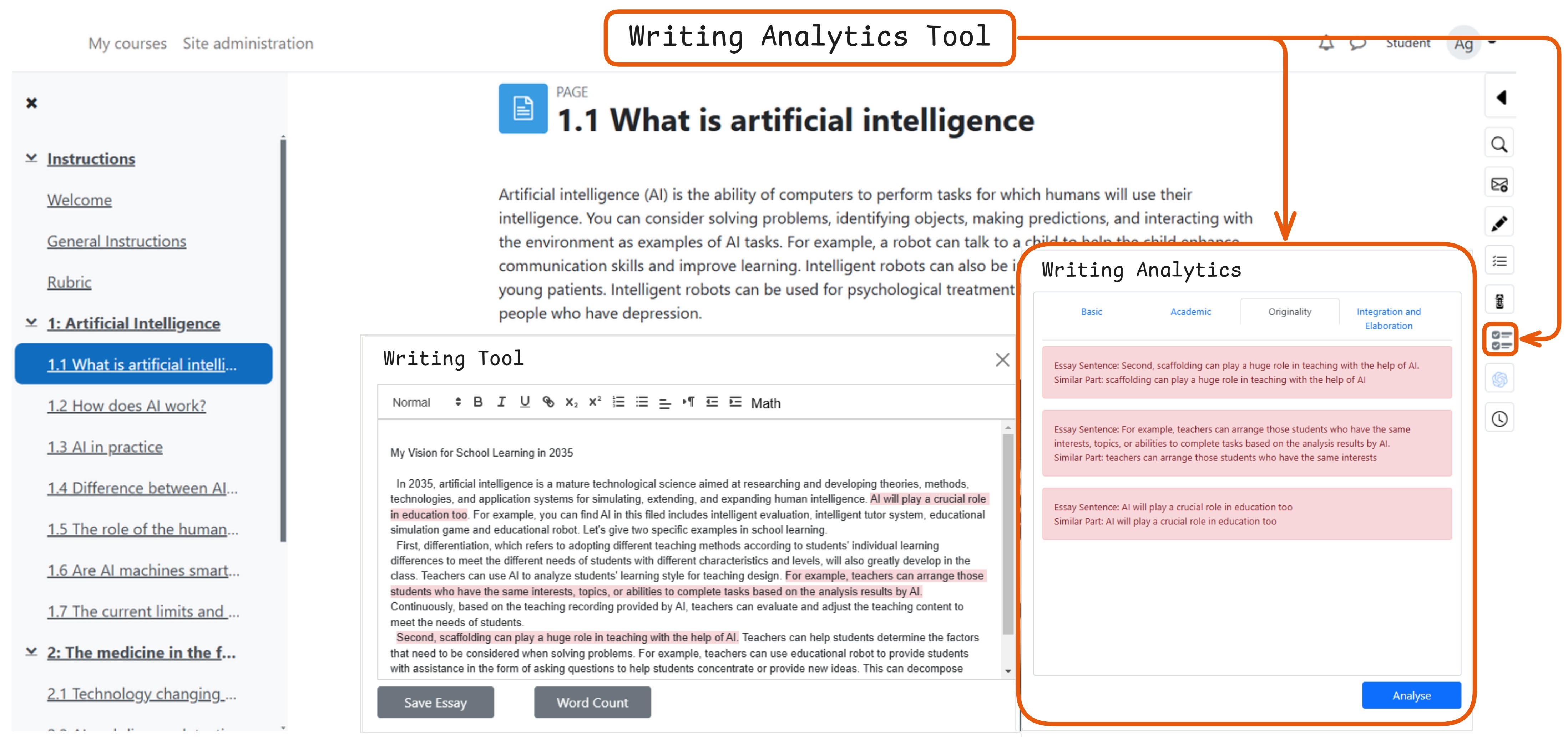}
\caption{Writing analytics tool - originality}
\label{fig:writing_analytics_tool_originality}
\end{figure}

The second function, the Academic Writing module (Figure~\ref{fig:writing_analytics_tool_vocab}), supports learners in aligning their language with academic conventions. This module draws on a back-end Key-Value based data structure and phrase-matching to efficiently identify issues with tone, sentence complexity, and vocabulary usage. By delivering context-specific guidance grounded in academic expectations, the tool promotes students' ability to plan and adapt their writing style to disciplinary norms, thus supporting SRL strategies related to goal-setting, strategic planning, and self-reflection

\begin{figure}[ht]
\centering
\includegraphics[width=0.85\linewidth]{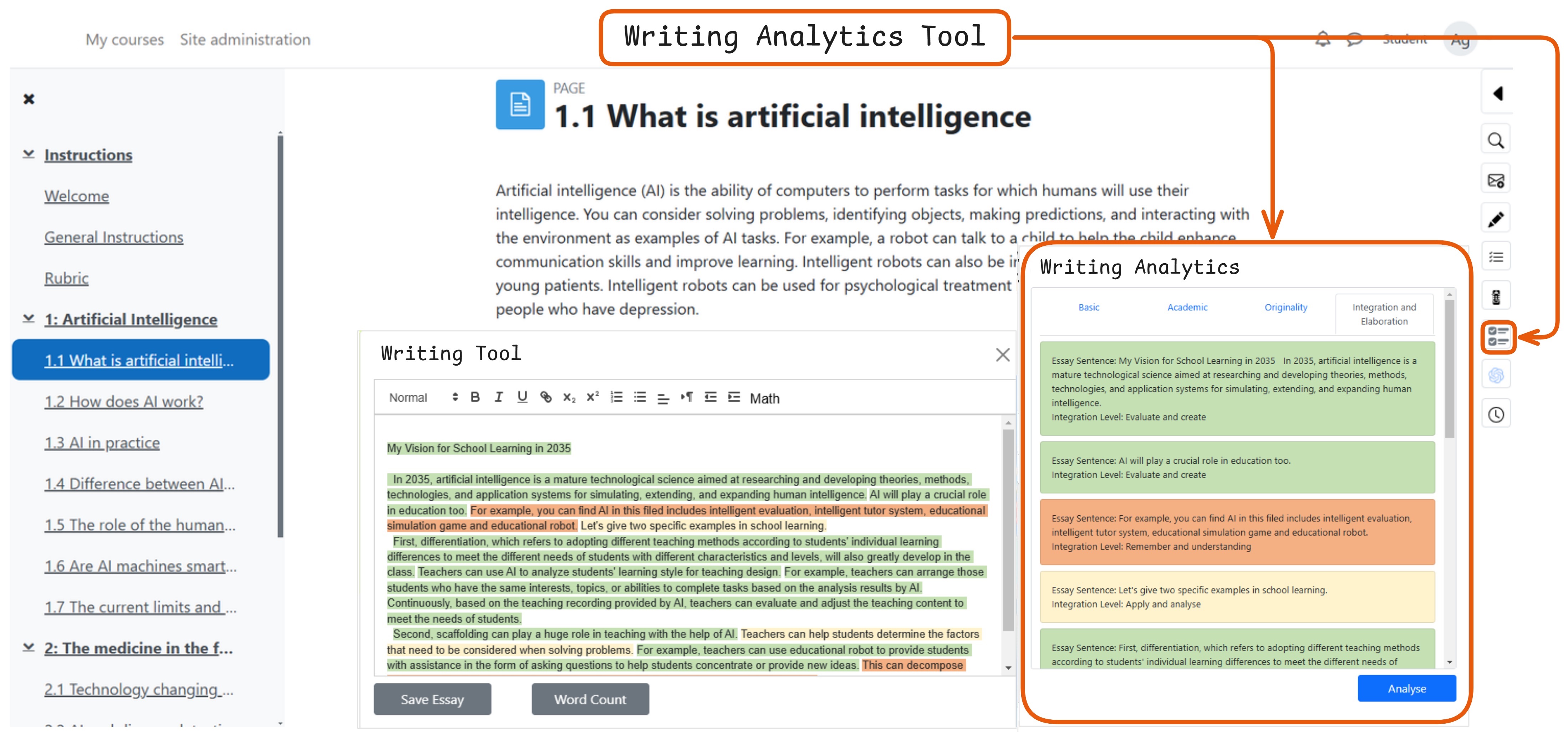}
\caption{Writing analytics tool - cognition classification}
\label{fig:writing_analytics_tool_cognition}
\end{figure}

The Originality function (Figure~\ref{fig:writing_analytics_tool_originality}) evaluates the uniqueness of student text by flagging any instance where more than seven consecutive words match the background reading material, leveraging an n-gram algorithm for detection. Beyond supporting academic integrity, the explicit highlighting of copied segments prompts students to monitor their use of sources, reconsider how they synthesise information, and actively paraphrase in their own words. This encourages the SRL processes of self-evaluation and strategic revision, empowering learners to internalise standards for originality.

Finally, the Cognition Classification module (Figure~\ref{fig:writing_analytics_tool_cognition}) analyses each sentence using Bloom’s taxonomy, classifying it according to cognitive level—ranging from remembering to creating—via the OpenAI davinci-002 model~\citep{forehand2010bloom,iqbal2023towards,iqbal2024towards}. By providing real-time feedback on the depth and complexity of their arguments, students are guided to elevate their essays through higher-order thinking. This function explicitly supports SRL by encouraging ongoing evaluation of cognitive engagement and targeted revision to achieve deeper analytical and synthetic writing.

Each module delivers immediate, transparent feedback by highlighting errors or cognitive levels within the student’s text. This direct correspondence between analytic results and source text empowers students to quickly identify strengths and weaknesses and strategically address them. For example, visible error markup in Basic and Academic Writing facilitates focused self-correction, while clear identification of copied passages in the Originality function scaffolds effective revision behaviour. The Integration and Elaboration tool’s feedback encourages learners to set higher goals for cognitive engagement, aligning their efforts with the objective of continuous improvement—a fundamental principle of SRL.

By integrating these targeted feedback mechanisms, the writing analytics tool not only enhances writing quality but also actively scaffolds students’ self-regulatory cycles of goal-setting, monitoring, and adaptive revision. Consequently, \flora’s analytics framework prepares learners to become independent, reflective writers capable of managing and advancing their own learning processes.

\subsection{Chatbot Support with Multiple GenAI-powered Agents}\label{subsec:chatbot}

With the rapid advancement of GenAI, its integration into educational contexts has become increasingly prevalent. GenAI-powered tools, such as chatbots, provide students with immediate access to information and personalised guidance, thereby offering significant potential to foster SRL. By facilitating processes such as setting learning objectives, selecting effective strategies, and reflecting on outcomes, GenAI tools can scaffold key stages of SRL. Within the \flora Engine, we have developed and integrated GenAI-based chatbot tools that leverage comprehensive learning contexts—including students’ prior knowledge, learning goals, and trace data logs—to systematically monitor progress and deliver real-time, actionable feedback. This continuous, contextualised support promotes metacognitive development, encouraging learners to reflect on their approaches and make adaptive adjustments throughout the learning process.

\begin{figure}[ht]
\centering
\includegraphics[width=0.85\linewidth]{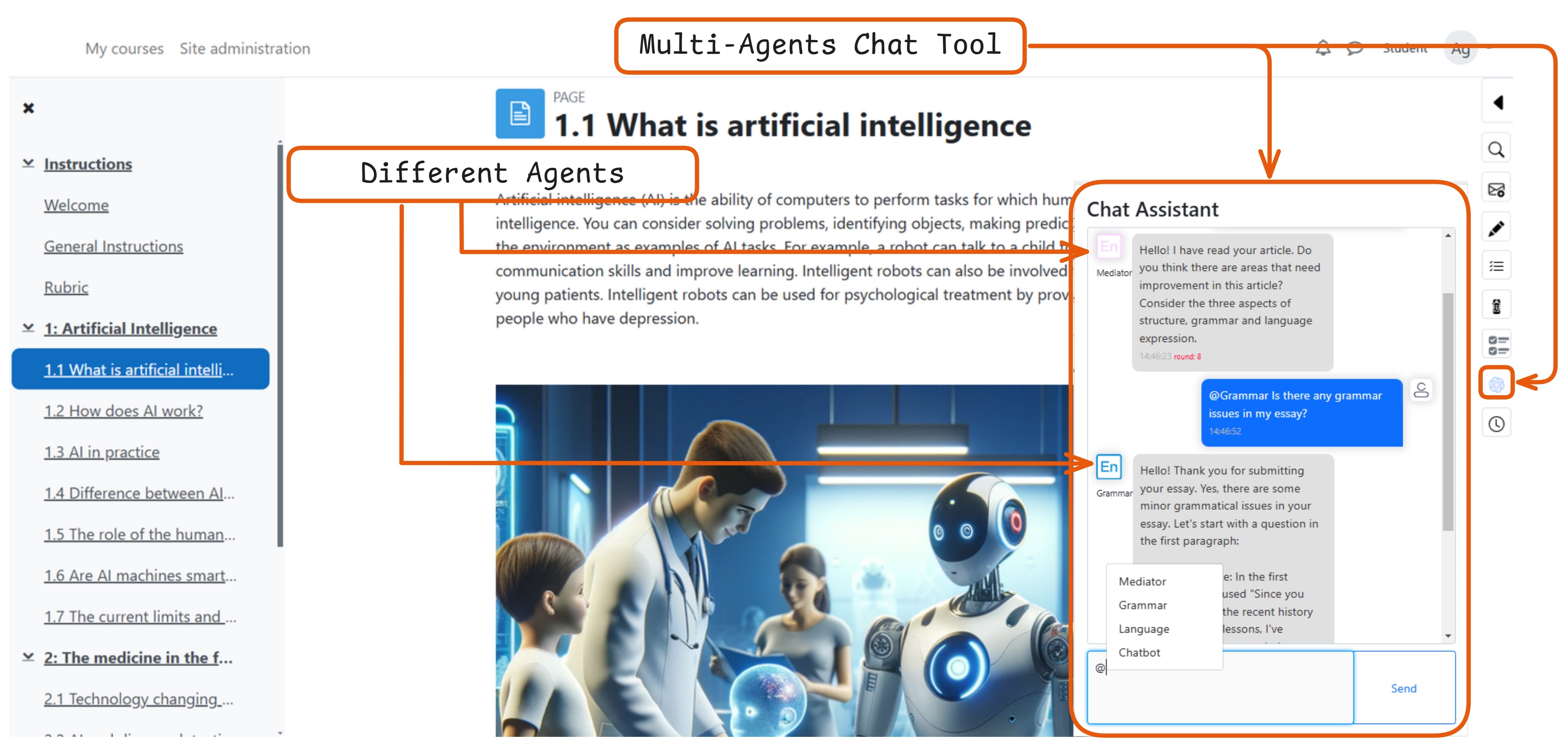}
\caption{Multi-Agents chatbot tool -- single chat window}
\label{fig:multi_agents_chatbot_tool_single}
\end{figure}

To accommodate diverse educational scenarios, the chatbot tool can be configured and used in flexible ways. The Figure~\ref{fig:multi_agents_chatbot_tool_single} employs a single chat window that supports multiple GenAI agents, each configured with distinct pre-prompts to specialise in specific domains or tasks. All agents share the conversation history, enabling learners to interact with a panel of experts within a cohesive dialogue. This collaborative environment mirrors authentic problem-solving situations, where engaging with multiple perspectives enhances the depth and breadth of feedback. By receiving collective input from specialised agents, students are better able to monitor their understanding, seek clarification, and receive nuanced guidance—thereby strengthening SRL components such as self-monitoring, strategy selection, and adaptive regulation.

\begin{figure}[ht]
\centering
\includegraphics[width=0.85\linewidth]{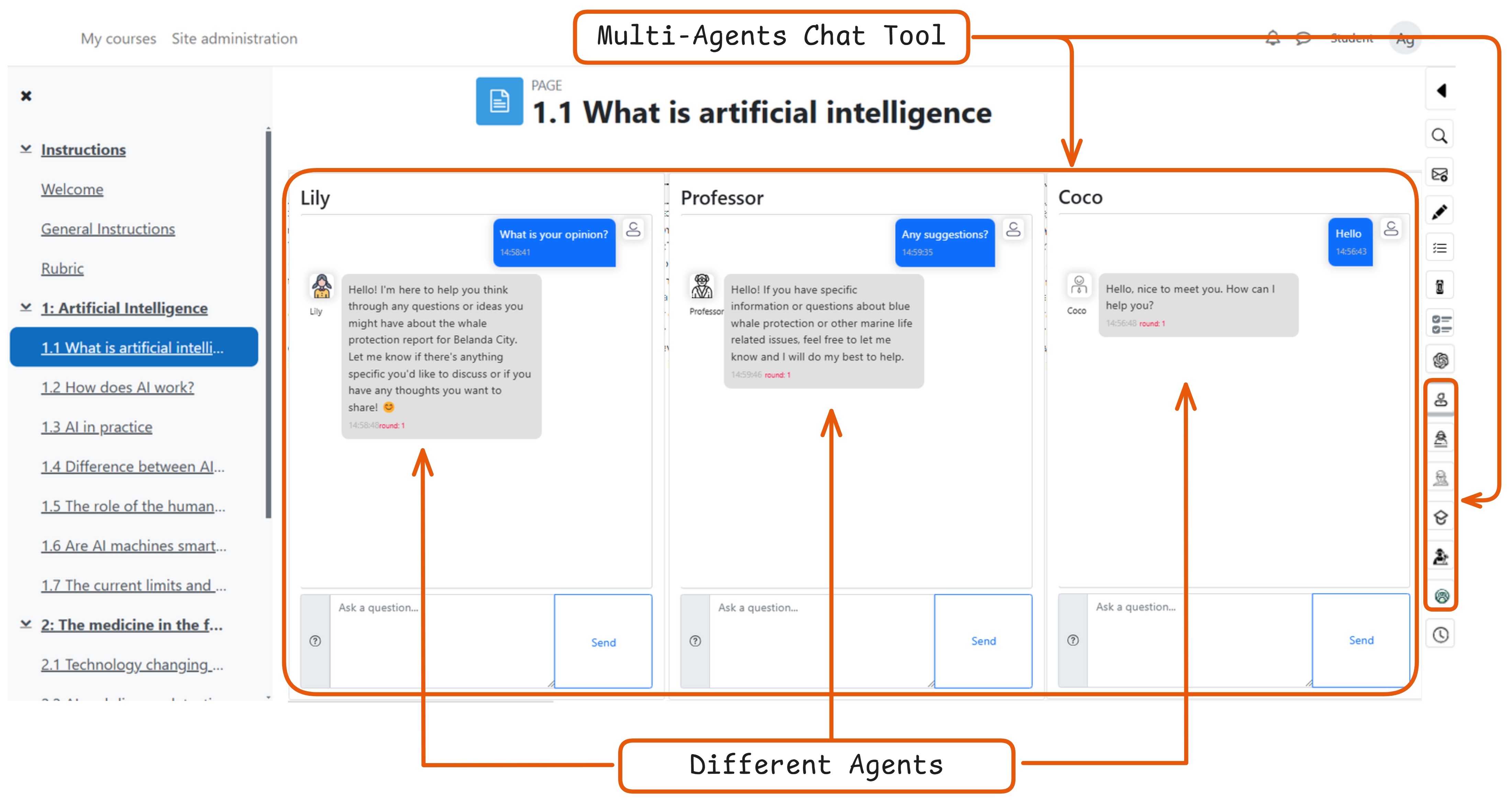}
\caption{Multi-Agents chatbot tool -- separated chat window}
\label{fig:multi_agents_chatbot_tool_multi}
\end{figure}

The Figure~\ref{fig:multi_agents_chatbot_tool_multi} adopts a multi-window configuration, assigning each GenAI agent to a separate, independent chat interface. In this setup, conversation histories are not shared, allowing each interaction to remain focused and distinct. This design is particularly effective in simulations or role-based scenarios where compartmentalised, parallel discussions are required. For example, in a negotiation exercise, one agent may serve as the client and another as the supplier, enabling learners to engage fully with each perspective. This separation supports SRL by encouraging learners to set targeted objectives, monitor progress within specific contexts, and evaluate outcomes independently for each scenario.

Both versions of the chatbot tool offer flexible configuration options, such as customizable pre-prompts and agent avatars, allowing educators to tailor the tool to specific course requirements and learning preferences. By seamlessly integrating these advanced GenAI chatbots into the \flora Engine, we provide learners with dynamic, interactive, and individually responsive support mechanisms. Ultimately, these tools enhance engagement, foster critical thinking, and promote the iterative self-assessment and strategy adjustment central to effective self-regulated and hybrid human-AI regulated learning.

\subsection{Collaborative Writing with Adaptive Support}\label{subsec:coll-writing}

\begin{figure}[ht]
\centering
\includegraphics[width=0.85\linewidth]{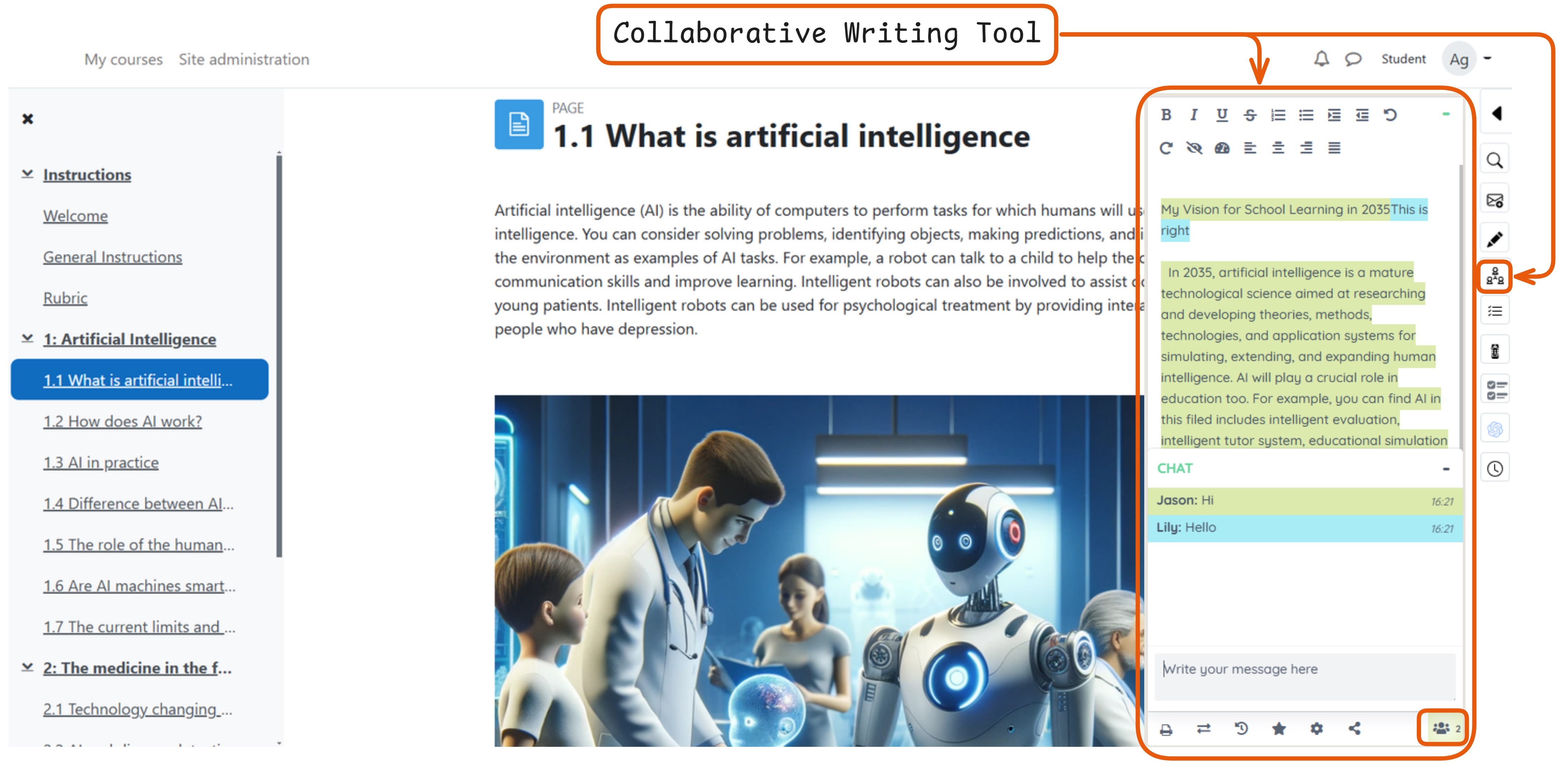}
\caption{Collaborative writing tool}
\label{fig:collaborative_writing_tool}
\end{figure}

To broaden the applicability of the \flora Engine, we have developed a collaborative writing tool designed to enhance both educational and research capabilities. This tool enables multiple users to simultaneously edit shared documents, similar to platforms like Google Docs (see Figure~\ref{fig:collaborative_writing_tool}). Distinctively, \flora’s collaborative tool records granular activity data—including keystrokes, revisions, and user interactions—in real time. This comprehensive data logging provides valuable insights into learners’ collaborative strategies, participation patterns, and group dynamics, thereby facilitating robust analysis of collaborative learning processes in authentic environments.

A key innovation of the \flora collaborative writing tool is the integration of a GenAI-powered chat interface. This embedded intelligent agent augments group communication by delivering context-sensitive feedback, answering queries, and generating relevant suggestions. The adaptive, personalised scaffolding provided by GenAI supports learners as they engage in real-time goal setting, monitor both their own and their peers’ contributions, and reflect on collaborative strategies. Consequently, students are empowered to exercise critical SRL skills such as self-monitoring, strategic revision, and adaptive planning, all within a supportive, interactive environment.

The fusion of AI and collaborative writing in \flora advances the concept of hybrid human–AI regulated learning. By mediating peer dialogue and offering individualised feedback, the GenAI agent fosters metacognitive development and the co-construction of knowledge. Learners actively regulate their learning by setting objectives, adjusting approaches, and evaluating outcomes, benefitting from immediate support tailored to both personal and group needs. This synergy enhances the depth and quality of collaboration, increases student engagement, and directly cultivates SRL competencies.

In addition to these innovative features, \flora encompasses a suite of other tools designed to support SRL across diverse educational contexts. Further details on these capabilities are provided in Appendix~\ref{sec:appendix_other_feature}.

\section{\flora Engine Implementation in Research Studies}
\label{sec:case-studies}

The \flora Engine has been adopted by more than 20 universities and institutions globally, supporting a substantial and diverse student population. This widespread adoption reflects its effectiveness in detecting, measuring, and facilitating SRL through advanced, research-informed tools for both educators and learners. As a result, the \flora Engine has become an integral resource for enhancing educational outcomes in varied instructional settings.

To illustrate its practical utility, efficacy, and adaptability, this section presents representative case studies from diverse educational contexts. These examples demonstrate how the \flora Engine supports a wide array of learning objectives and tasks, showcasing its versatility and positive impact on learning processes. Through an examination of its application across scenarios, we offer evidence for the Engine's capacity to advance SRL and HHAIRL in authentic environments. Because the system can be configured in different ways across contexts, the case studies adopt different research objectives, data structures, and analytical approaches. For this reason, statistical reporting is not fully homogeneous; instead, it is adapted to the methodological requirements of each case.

\subsection{Case 1 -- Comparing Human, Rule-Based, and GenAI Feedback}
The first study investigated the effectiveness of different support mechanisms in enhancing English writing skills among university students who are English as a Second Language (ESL) learners. The primary objective was to assess how various forms of assistance—including GenAI, human expert guidance, and writing analytics tools—impact students' performance in a two-stage English reading and writing task.

\textbf{Participants:} A total of 117 university students participated in the study conducted from July to September 2023. The participants had an average age of 22.61 years (SD = 3.39), with 70\% identifying as female and 55\% being undergraduates. They are from diverse disciplines and shared a common first language other than English. All were ESL learners.

\begin{figure}
    \centering
    \includegraphics[width=0.9\linewidth]{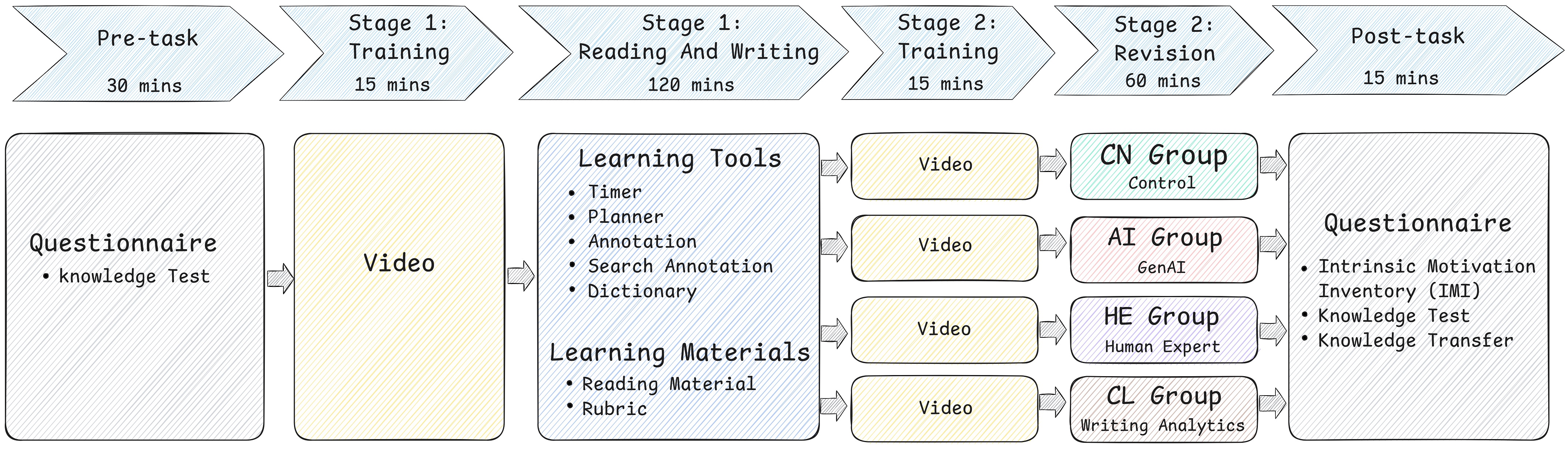}
    \caption{Experimental procedure of case study 1}
    \label{fig:Experimental_case_1}
\end{figure}

\textbf{Research Procedure:} The study was conducted in a controlled laboratory setting and followed a six-step procedure, as illustrated in Figure \ref{fig:Experimental_case_1}: pre-task questionnaire, Stage 1 training about the use of \flora, Stage 1 reading and essay writing task, Stage 2 training about the use of \flora, Stage 2 essay revising, and post-task questionnaire. Participants used computers to complete questionnaires, watch training videos, and engage in the learning tasks. In Stage 1, after watching a training video on how to use the learning tools provided in the learning environment, participants undertook a two-hour reading and writing task. They were required to read materials on three topics — artificial intelligence, differentiated teaching, and scaffolding teaching — and write an essay envisioning the future of education in 2035 by integrating these topics. A rubric outlining the grading criteria was provided for reference during the writing process. In Stage 2, participants watched a second training video introducing the specific support mechanisms available to their assigned experimental group. They then engaged in a one-hour revision task, aiming to improve their essays using the provided support. After completing the tasks, participants were asked to complete a post-test within one day.

Participants were randomly assigned to one of four experimental groups, each receiving distinct forms of support during the revision stage. The Control Group (CN, n = 30) received no additional support, completing the revision independently and experiencing no changes to the learning environment across both stages. The AI Group (n = 35) received assistance from a GPT-4-based chatbot tool (see Section~\ref{subsec:chatbot}), embedded within the user interface. This AI support was limited to task-related content through targeted pre-prompts, ensuring conversations remained relevant to the learning objectives. The Human Expert Group (HE, n = 25) received guidance from a qualified academic writing specialist. Participants in this group could interact with the expert via an instant chat tool (see Appendix~\ref{sec:appendix_other_feature}) to refine and enhance their essays. The Writing Analytics Tools Group (CL, n = 27) utilised a writing analytics toolkit (see Section~\ref{subsec:writing-analytics}), which provided automated feedback on spelling and grammar, academic style, originality, and cognition classification. 

\textbf{Key Research Findings:} This study yielded several key findings regarding learners’ interactions with GenAI-based chatbots, human experts, and writing analytics tools. First, first-order Markov models (FOMM), used as a process mining technique, were used to analyse help-seeking stages and activities within each identified process. These stages included \textit{Diagnosing Question}, \textit{Asking Help}, \textit{Evaluating Help}, and \textit{Processing Help}. The process mining results showed that learners who engaged more with GenAI-based chatbots often bypassed \textit{Diagnosing Question} or \textit{Evaluating Help} which are related to metacognition~\citep{chen2025unpacking}. The Mann-Whitney U test results indicated a significant difference between the two groups in the \textit{Asking Help} stage ratio ($Z = -3.334, ES = 1.288, p < 0.001$), with the AI Group showing a significantly higher ratio than that of HE Group. This is because GenAI-based chatbot provided instantly accessible resources, which naturally encourage more executive and direct help-seeking behaviours. This divergence in questioning behaviours mediated the relationship between support type and academic performance: learners who asked direct questions—typically supported by GenAI—showed greater improvements in essay revisions due to efficient knowledge acquisition~\citep{cheng2025asking}. However, reliance on GenAI sometimes reduced metacognitive engagement, as learners tended to delegate metacognitive monitoring to the chatbot, potentially limiting deep learning and critical reflection~\citep{chen2025unpacking}. Second, after the experiment, 59 participants agreed to participate in the semi-structured interviews (33 in the AI group and 26 in the HE group, lasting 35 minutes). The interview transcription data was analysed using thematic analysis assisted by NVivo 12. The results showed that the chatbot fostered social comfort and autonomy by minimizing social judgment, but was limited in responsiveness and raised privacy concerns. Conversely, human experts provided personalised and responsive support but occasionally constrained learner autonomy. Learners’ value priorities varied based on personal preferences and contexts, resulting in inherent trade-offs in their learning experiences~\citep{shen2025aligning}. Third, in another study, learners' motivation was measured using the Intrinsic Motivation Inventory~\citep{mcauley1989psychometric, torbergsen2023nursing} in the post-task to compare intrinsic motivation across groups on four subscales: Interest/Enjoyment, Perceived Competence, Effort/Importance, Pressure/Tension. One-way ANOVAs followed by Tukey’s HSD were performed for each subscale. No significant between-group differences were observed for Interest/Enjoyment ($F = 1.087, p = 0.358, \eta^2 = 0.029$), Perceived Competence ($F = 0.453, p = 0.716, \eta^2 = 0.012$), Effort/Importance ($F = 1.152, p = 0.332, \eta^2 = 0.030$), or Pressure/Tension ($F = 0.546, p = 0.652, \eta^2 = 0.015$). Complementary analyses of SRL process frequencies and process mining indicated that the GenAI condition involved fewer metacognitive activities. Thus, although the GenAI chatbot often yielded immediate performance gains (e.g., higher essay scores), these gains were not accompanied by increases in intrinsic motivation and did not generalise to longer-term knowledge outcomes relative to human tutoring or independent study. The convenience of GenAI support at times led to "metacognitive laziness", with learners engaging less in planning and monitoring, both vital for SRL~\citep{fan2025beware}. Fourth, the design of formative writing analytics tools significantly influenced learners’ self-assessment processes. An epistemic network analysis~\citep{shaffer2016tutorial} was conducted to examine the impact of different feedback tools on learners' writing outcomes. The results demonstrated that tools offering high-affordance feedback, such as clear grammatical corrections, supported accurate self-assessment and substantive improvement in writing. Conversely, low-affordance tools, including those integrating Bloom’s taxonomy classifications, often hindered effective self-reflection and revision due to lack of actionable guidance or potential for confusion~\citep{tang2024facilitating, forehand2010bloom}. Finally, although learners generally preferred human feedback, preferences were shaped by the type of GenAI tool used. High-performance, interactive chatbots like ChatGPT (see Section~\ref{subsec:chatbot}) fostered stronger preferences for AI-supported learning, whereas less interactive or rule-based tools (see Section~\ref{subsec:writing-analytics} and Appendix~\ref{sec:appendix_other_feature}) led learners to favour human tutors for their greater personalization and interactivity~\citep{le2025rolling, cheng2025asking}.

\subsection{Case 2 -- Clinical Reasoning Training}

This study explores the effectiveness of using GenAI-based chatbot tool, which is adjusted with preprompt to simulate as a Virtual Standard Patient (VSP), to support medical students' SRL and HHAIRL, with the ultimate goal of enhancing clinical reasoning skills. By simulating realistic patient interactions, the research aims to assess whether engaging with GenAI VSP improves students' history-taking abilities, diagnostic reasoning, and overall clinical competence. Additionally, it examines the impact of a GenAI-based virtual teaching assistant on providing personalised scaffolding to support learning.

\textbf{Participants:} A total of 210 second-year clinical medical students from a Medical university participated in this study. All participants had completed foundational training in symptomatology and history-taking. However, many self-reported their clinical reasoning abilities as average, highlighting the need for additional skill development in this area.

\begin{figure}
    \centering
    \includegraphics[width=0.9\linewidth]{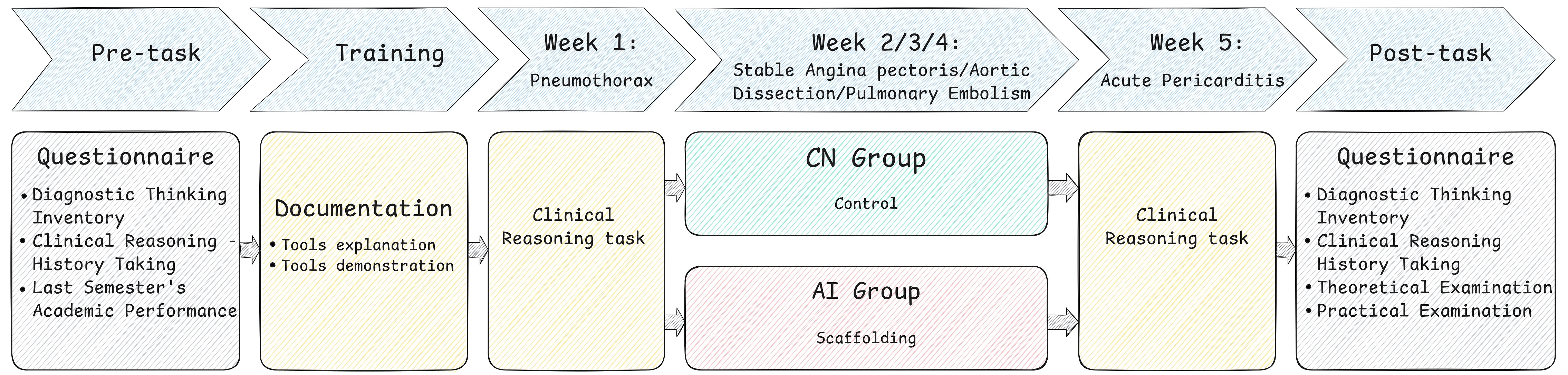}
    \caption{Experimental procedure of case study 2}
    \label{fig:Experimental_case_2}
\end{figure}

\textbf{Research Procedure:} The study consisted of five stages, illustrated in Figure~\ref{fig:Experimental_case_2}: a pre-task survey, a training session, the main task (comprising five consecutive clinical reasoning tasks), and a post-task survey. In the pre-task survey, participants provided demographic data, previous course grades, and self-assessments of their history-taking and clinical reasoning abilities. The training session required participants to review task instructions and view a demonstration video on the \flora learning environment. The core of the study involved five weekly clinical reasoning tasks, each focused on a distinct medical topic including Spontaneous Pneumothorax, Stable Angina, Aortic Dissection, Acute Pulmonary Embolism, and Acute Pericarditis. Each task lasted 120 minutes. During these sessions, students interacted with the VSP via a chat interface to gather information and formulate a diagnostic conclusion. They were able to review instructions, interact with the VSP, take notes, draft, and submit diagnostic conclusions. Upon submission, students received automated GenAI-based feedback from the \flora system according to predefined rubrics (see Section~\ref{subsec:chatbot}). After completing the main tasks, participants completed a post-task survey to reassess their skills and reflect on their learning experience.

Participants were randomly assigned to either the GenAI Support group (AI Group, n = 106) or the control group (CN Group, n = 104). During modules 2–4 of the main task, the AI group interacted with both the GenAI VSP and a GenAI teaching assistant. This teaching assistant provided individualised feedback, guidance, and adaptive scaffolding to support learning. In contrast, the control group engaged only with the GenAI VSP, allowing for a direct comparison of the effects of personalised GenAI support. Both groups accessed the \flora platform’s suite of tools, which included the GenAI VSP chat interface powered by ChatGPT models, an auto-saving note-taking tool with data collection features, a text submission tool for diagnostic results that offered immediate feedback and performance scores (see Appendix~\ref{sec:appendix_other_feature}), and a timer to regulate task duration.

\textbf{Key Research Findings:} The study assessed students' clinical reasoning skills by analysing their history-taking behaviours during interactions with VSPs. By analysing 1,030 consultation dialogues, a Pearson correlation analysis was conducted, and students' dialogues were coded into 12 categories. Some behaviours showed strong correlations (r = 0.208–0.225, \(\chi^2\)=6.019, all \textit{P} < .05) with reasoning metrics like diagnostic accuracy, history-taking scores, and clinical knowledge test results. This is because these behaviours reflect essential components of effective clinical reasoning: organizing questions in a logical sequence aids in systematic data collection; asking specific, pathophysiologically relevant questions demonstrates a deeper understanding of disease processes; and summarising and reorganising information indicates the ability to synthesise patient data. These behaviours enhance students' ability to accurately diagnose and understand clinical scenarios, thereby improving their performance on reasoning metrics. By integrating GenAI-driven VSPs into the \flora platform, the study offered scalable and interactive opportunities for students to practice history-taking in a realistic, simulated environment. This approach enabled dynamic patient consultations and provided immediate, personalised feedback, thereby enhancing students’ clinical reasoning skills~\citep{chen2025medicalcoding}. However, GenAI can be prone to generating hallucinations, resulting in the presentation of inaccurate or fabricated information~\citep{qian2025towards}. Consequently, the GenAI-based VSPs may sometimes provide students with misleading responses. This can hinder students’ diagnostic reasoning by introducing errors, and thus raises important concerns about the reliability of using GenAI in medical education.

Beyond generating valuable insights in medical diagnostics, this study also demonstrates the versatility of the \flora platform for a wide range of experiential learning tasks. Its capacity to simulate interactive scenarios suggests potential applications in leadership training and other domains requiring experiential learning. The results highlight that integrating \flora with GenAI technologies can facilitate skill development and deliver personalised learning experiences across diverse educational contexts. And, because of the personalised and adaptive support throughout the learning process, thereby the development of SRL skills can also be facilitated.

\subsection{Case 3 -- advancing analytics-based adaptive SRL scaffolding with GenAI}
\begin{figure}
    \centering
    \includegraphics[width=0.8\linewidth]{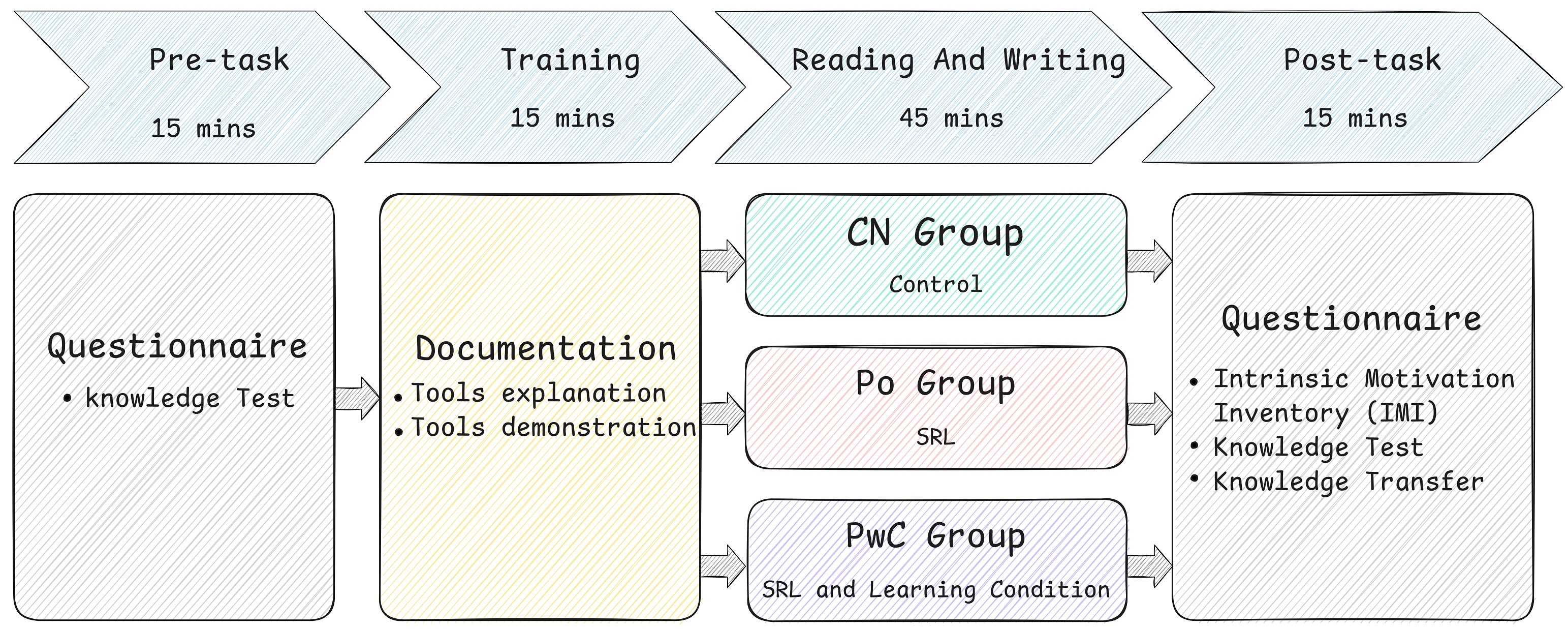}
    \caption{Experimental procedure of case study 3}
    \label{fig:Experimental_case_3}
\end{figure}

This project investigated learners’ SRL skills in both K–12 and higher education contexts, with a focus on how learning analytics and AI can be leveraged to support robust SRL processes in the era of AI. A central innovation of this work is the integration of learning analytics with GenAI to enhance the adaptivity of SRL scaffolding. Prior research has identified the limitations of rule-based analytics, which often fail to accommodate the unique strategies and conditions of individual learners~\citep{li2025turning}. To overcome these limitations, this project utilised real-time learning analytics to inform GenAI models, enabling the generation of personalised and contextually relevant SRL support.

\textbf{Participants:} Data were collected from both secondary and higher education settings. In the K–12 context, data were gathered from 663 students across 14 schools in six countries: Brazil (86 students from one school), Colombia (33 from one school), India (93 from three schools), Australia (88 from three schools), China (213 from two schools), and the United Arab Emirates (150 from four schools). In higher education, a reflective writing task was implemented as part of an academic English writing course, with data collected from 171 undergraduate students.

\textbf{Research Procedure:} This project comprised four sequential learning modules. The first module, the pre-task phase, included three activities: a demographic questionnaire (collecting data such as gender, age, and first language), a domain knowledge assessment on AI applications in medicine (15 multiple-choice questions), and a learning strategy knowledge test based on the ISDIMU instrument \citep{bannert2021isdimu}. In the latter, participants responded to hypothetical scenarios by indicating the likelihood of employing specific learning strategies; for example, identifying whether they would consult solution materials before attempting a new task.

Following the pre-task module, students entered the training phase, which was designed to familiarise them with the learning platform and its built-in tools. Instructional materials were provided in a text-and-image format to support user orientation.

The third module, the core reading and writing task, required students to compose a 200–300 word essay using information from multiple texts about AI applications in medicine. During this task, students utilised the \flora learning tools and were randomly assigned to one of three groups (see Figure~\ref{fig:Experimental_case_3}). The control (CN) group completed the essay with no explicit SRL support. The Process-only (Po) group received GenAI-based scaffolds tailored to their real-time SRL processes, which were identified by analysing prior task data using time-series mapping and rain-cloud plots at seven-minute intervals \citep{van2023design}. For each interval, SRL processes exhibiting significant differences in frequency between high and low performers were selected to inform the scaffolding in the Po group. Scaffolding was triggered if critical SRL processes—such as monitoring instructions or rubric use (common among high performers)—were absent within predetermined timeframes. For example, if a learner did not review the task instructions within 14 minutes, a scaffold prompted this behaviour; if the student had already done so, the scaffold was not activated.

The Process with Condition (PwC) group received similar GenAI-generated process-based scaffolding, but with content further personalised by individual learning conditions derived from the COPES model of SRL \citep{winne1998studying, winne2022modeling}. These conditions included factors such as domain knowledge and awareness of available resources. Learning analytics techniques were used to detect these real-time conditions, allowing GenAI to provide adaptive, individualised scaffolding. Details of the scaffolding mechanisms and design are provided in Section~\ref{subsec:GenAI-scaffolds}.

Upon completing the main writing task, students proceeded to the post-task module, which featured two activities: a post-test using the same 15 multiple-choice questions to assess learning gains, and a feedback literacy survey evaluating attitudes, mindsets, and behaviours related to feedback \citep{Dawson2023}.

\textbf{Key Research Findings:} The design of this project has been implemented in empirical studies across both K-12 and higher education settings. In the higher education context, \cite{li2025turning} conducted a randomised controlled trial compared the three groups as described above: CN, Po and PwC groups. Using Ordered Network Analysis (ONA), learners' SRL processes—and how those processes transition among one another—onto a two-dimensional space. Through a means-rotation method, differences were maximised on the first dimension, and a Wilcoxon test was subsequently performed. The results indicated that learners in the PwC group demonstrated significantly more effective and metacognitive SRL behaviors, such as frequently consulting task instructions and rubrics, compared to the other groups (\emph{PwC vs.\ CN}: $W$ = 1944, $n$ = 112 , $p$ = 0.0213, $r$ = 0.707; \emph{PwC vs.\ Po}: $W$ = 1149, $n$ = 109, $p$ = 0.0478, $r$ = 0.863). Furthermore, the study employed Dynamic Time Warping (DTW) with k k-means clustering to assess learners' compliance (i.e., the level of adherence to the suggested processes) with the SRL scaffolds, and differences between compliers and non-compliers were mapped onto the ONA space. The results revealed that learners' compliance varied significantly: compliers demonstrated clear adherence to scaffold recommendations and engaged in more frequent metacognitive processes, particularly during writing phases. In contrast, non-compliers exhibited a more linear, less strategic learning trajectory ($W$ = 596, $n$ = 54, $p$ < 0.000, $r$ = 0.548). These findings highlight the importance of incorporating both learners’ real-time SRL processes and evolving conditions in scaffold design and provide empirical evidence supporting the scalability and effectiveness of GenAI-powered adaptive scaffolding for enhancing SRL in computer-based learning environments. By using this approach, our tool demonstrates the capability of delivering personalised support that adapts not only to learners' real-time SRL processes but also to dynamic learning conditions that shift throughout the task. This adaptivity, in turn, promotes improved compliance and long-term impacts in facilitating effective self-regulation in the age of AI.

However, it is important to note that these findings did not reveal a direct effect on learners' writing performance, as measured by their scores. This outcome may be attributed to the tools’ primary focus on supporting SRL activities, which are intended to foster long-term learning development rather than immediate performance gains. Such an interpretation aligns with prior research suggesting that SRL interventions typically do not produce immediate or significant improvements in essay performance~\citep{yan2024promises}. In another study, \cite{jin2025analytics} explored the relationship between feedback literacy and SRL processes among secondary school students across different countries. Through K-medoids clustering of self-reported feedback literacy, the study identified two distinct clusters: Proactive Feedback Engagers and Moderate Feedback Engagers. ONA revealed that Proactive Feedback Engagers initially employed less effective learning strategies but adapted significantly following scaffolded feedback, demonstrating more balanced and strategic regulation. In contrast, Moderate Feedback Engagers began with more strategic initial approaches but showed less adaptability and responsiveness to scaffolding suggestions. These findings emphasise the dynamic, context-specific nature of feedback literacy and its interplay with SRL, highlighting the need for adaptive scaffolding tailored to students’ feedback literacy profiles to enhance engagement and SRL processes.

\section{Discussion}\label{sec12}

The \flora Engine represents a significant advancement in the development of AI-enhanced educational platforms, focusing on addressing the challenges and opportunities of hybrid Human-AI regulation in SRL. This study provides evidence of \flora’s ability to both facilitate SRL processes across various contexts and domains and contribute to the broader understanding of how learners interact with AI systems within complex learning environments. 

\subsection{Advancing Hybrid Human-AI Regulation in Education}

\flora demonstrates the potential of GenAI and learning analytics to advance hybrid Human-AI regulation by supporting and scaffolding learners’ SRL behaviours. The platform detects critical SRL processes, assesses real-time learning conditions and, when necessary, intervenes via personalised scaffolds designed to enhance students’ cognitive, metacognitive, and motivational regulation. This aligns closely with emerging frameworks such as Hybrid Human-AI Regulated Learning (HHAIRL) \citep{huang2024promoting,molenaar2022concept}, which emphasise the synergetic interplay of human agency and AI augmentations in fostering learner autonomy. For instance, the scaffolding intervention mechanisms triggered by unmet SRL benchmarks (e.g., the absence of critical planning or monitoring behaviours during timed tasks) illustrate how AI can enhance learners’ self-regulatory processes while maintaining their central role in decision-making and goal-setting.

Across all three case studies described in Section \ref{sec:case-studies}, the results suggest that personalised scaffolding, particularly when grounded in dynamic trace data, can promote learners’ active engagement in SRL behaviours such as goal setting, process monitoring, and self-evaluation. A notable example is the integration of GenAI personalisation in writing revision tasks, where learners in the AI-backed scaffolding groups exhibited improved essay scores, a finding consistent with prior work showing how adaptive scaffolding fosters task-relevant SRL behaviours \citep{jarvela2023human, edwards2025human}. These findings contribute to the broader literature on co-regulated SRL approaches by revealing how trace analytics combined with GenAI can dynamically adjust scaffolds to learners’ changing conditions, a capability absent in traditional rule-based systems \citep{huang2024promoting}.

However, the findings also highlight the risks of over-reliance on AI systems, echoing concerns raised in prior studies~\citep{fan2025beware}. For instance, the case study of ESL learners using GenAI-based support for writing revealed instances of "metacognitive laziness," where learners offloaded planning and monitoring to AI systems. While this improved task-specific outcomes in the short term, it may hinder the long-term development of learners’ critical SRL abilities, a finding corroborated by research on the pitfalls of AI augmentation in learning contexts \citep{molenaar2022concept}. 

\subsection{Fostering Collaboration and Personalisation Across Contexts}

The adoption of \flora across diverse educational domains, from reflective writing tasks to clinical reasoning and K-12 classroom settings, highlights its flexibility and scalability. Each case study shed light on the nuanced needs of learners within specific contexts, further emphasising the importance of adapting tools like \flora to align with disciplinary requirements. For instance, the use of the GenAI-based Virtual Standard Patient in the medical diagnostic reasoning task underscores the platform’s capability to simulate real-world problem-solving environments. The tool effectively supported learners’ organisation of diagnostic queries and improved history-taking practices, reflecting the potential for hybrid regulation to complement domain-specific pedagogical designs \citep{chen2025medicalcoding}.

The findings from Case 3 further highlight the importance of delivering personalised scaffolds not solely based on learners' SRL processes, but also aligned with individual attributes such as learner profiles and literacy levels (e.g., feedback literacy). While prior research has primarily focused on identifying and responding to learners' real-time SRL behaviours \citep{winne2017learning, viberg2020self}, our study extends this perspective by demonstrating how scaffold effectiveness increases when these supports are tailored explicitly to learner characteristics. For instance, learners' differing levels of feedback literacy considerably influenced their receptiveness to scaffolded interventions, underscoring the need for careful alignment between scaffold design and learners' individual profiles. GenAI offers a distinct advantage over traditional rule-based systems in this context, as it can dynamically incorporate varied learner attributes and literacy levels into scaffold design, enabling deeper personalisation beyond what is feasible through predefined logic alone. This finding supports recent theoretical arguments calling for comprehensive, learner-centred approaches in adaptive scaffolding designs within hybrid Human-AI regulated learning environments \citep{jarvela2023human, cukurova2024interplay, nguyen2025human}.

\subsection{Implications for Future Design and Research}

Firstly, providing learners with direct access to visual analytics presents a promising avenue for future development. Learner-facing dashboards that depict real-time SRL activities can enhance metacognitive monitoring, prompting students to reflect and make targeted adjustments to their learning strategies~\citep{li2023analytics}. Such tools can help cultivate meta-level awareness and encourage the gradual shift from external scaffold reliance to self-directed regulation.

Secondly, integrating Bloom’s taxonomy into \flora’s writing analytics demonstrated its capacity to improve the accuracy of students’ self-assessment and metacognitive monitoring \citep{tang2024facilitating}. However, effective use of formative feedback based on cognitive domains requires preliminary learner training \citep{forehand2010bloom}. Expanding \flora to include educator training modules, which complement learner-focused SRL interventions with teacher-led scaffolding, could further enhance the platform’s ability to support a diverse user base.

Finally, as highlighted in Case 3, adaptive scaffolding should consider both learners’ SRL processes and their individual characteristics, including domain knowledge and feedback literacy. Future iterations of \flora should leverage GenAI’s capacity to dynamically integrate real-time learner conditions, potentially using multimodal data such as eye-tracking for attention and facial expressions for confusion. This level of personalization surpasses traditional rule-based systems and supports recent theoretical calls for holistic, learner-centred analytics frameworks \citep{cukurova2024interplay, molenaar2022towards}, ultimately advancing hybrid human–AI regulated learning environments.

\subsection{Limitations of the Study and Future Directions}

While the results demonstrate \flora’s capacity to support hybrid Human-AI regulation and advance SRL-focused educational research, this study has certain limitations. First, the reliance on digital trace data restricts visibility into higher-order cognitive and emotional SRL processes such as self-motivation, frustration management, or contextual sense-making \citep{saint2020combining}. Incorporating multi-modal data sources, such as physiological signals or speech analysis, may enhance the comprehensiveness of SRL detection, especially motivational and affective aspects \citep{rakovic2024measuring}, in future development and research of \flora. Second, the experimental tasks varied significantly in domain and structure (e.g., reflective writing versus clinical reasoning), which may partially confound comparisons of scaffolding efficacy across contexts. Future studies could standardise task designs to isolate the effect of scaffolding interventions under more controlled conditions. Furthermore, longitudinal studies are needed to examine the lasting impact of hybrid Human-AI regulation on learners’ metacognitive engagement and SRL skill development over time. 

The ethical implications of embedding GenAI systems into personalised learning contexts remain under-explored. For example, anonymity, privacy concerns, and data use transparency should be more systematically addressed in both the research design and platform development stages to ensure compliance with ethical standards in diverse regions and populations. In addition, safeguards such as privacy-preserving data processing and informed consent mechanisms should be embedded into the platform design to ensure transparency and learner trust. Beyond ethics, the risk of learner over-reliance on AI support warrants attention: novice learners may risk becoming dependent on external scaffolds, while more advanced learners may benefit from complementary support~\citep{koedinger2007exploring,belland2017synthesizing}. Future research should therefore investigate how to calibrate scaffold intensity and adapt interventions to learner needs, balancing guidance with opportunities for autonomy and fostering long-term SRL skill development~\citep{molenaar2022concept}.

\section{Conclusion}\label{sec13}



The enhanced \flora Engine represents a significant step forward in supporting Hybrid Human-AI Regulated Learning by providing robust research tools and a flexible, open-source infrastructure for further development. Through the integration of GenAI-driven instrumentation tools, a real-time trace parser, and adaptive scaffolding, \flora enables detailed exploration of SRL processes and the complex dynamics between learners and AI. Its ability to capture and visualize real-time data facilitates the comparison of learning patterns with and without AI involvement, yielding valuable insights into the potential and limitations of AI in SRL.

However, it is important to acknowledge that the implementation of GenAI tools, while promising for enhancing immediate learner performance, is not without risks. If employed without careful pedagogical guidance, such interventions may inadvertently lessen learners’ metacognitive engagement and autonomy~\citep{darvishi2024impact}. Therefore, the use of \flora must be informed by explicit instructional strategies that promote reflective and critical engagement with AI, rather than reliance on automated support alone.

From a practical standpoint, \flora offers significant implications for various stakeholders. Educators can leverage \flora to design targeted SRL interventions and to monitor learner progress with greater granularity. Educational technology developers may benefit from its open-source framework, which allows for customization and integration with diverse learning environments. Institutional leaders, meanwhile, can utilize the system’s analytics to inform policy decisions and resource allocation in support of evidence-based, AI-enhanced education.

Looking ahead, several avenues for future research and development merit consideration. Expanding \flora’s capabilities to incorporate multimodal data—such as voice, gesture, or physiological signals—could yield a more comprehensive understanding of SRL processes and learner-AI interactions. Additionally, enhancing the explainability of AI-generated recommendations within \flora is essential for fostering transparency and trust among both learners and educators. Finally, dedicated training for teachers in the critical and effective use of GenAI tools will be crucial to maximize educational benefits while mitigating potential shortcomings.

In conclusion, by bridging the gap between automated AI-driven support and human agency, \flora contributes meaningfully to the development of HHAIRL research. Nevertheless, its effectiveness is contingent on thoughtful pedagogical integration, ongoing research, and proactive stakeholder engagement. Such a balanced approach will ensure that advances in AI not only empower learners to regulate their own learning but also enrich our collective understanding and practice of human-AI collaboration in education.









\section*{Declaration of Competing Interest}
The authors declare that they have no known competing financial interests or personal relationships that could have appeared to influence the work reported in this paper.

\section*{Declaration of generative AI and AI-assisted technologies in the writing process}

During the preparation of this work the author(s) used ChatGPT in order to check grammar of the paper. After using this tool/service, the author(s) reviewed and edited the content as needed and take(s) full responsibility for the content of the publication.


\printcredits

\bibliographystyle{cas-model2-names}

\bibliography{reference}
\appendix
\FloatBarrier
\section{Appendix - Other Features}\label{sec:appendix_other_feature}

\subsection{Other Tools}

\begin{figure}[ht]
\centering
\includegraphics[width=0.85\linewidth]{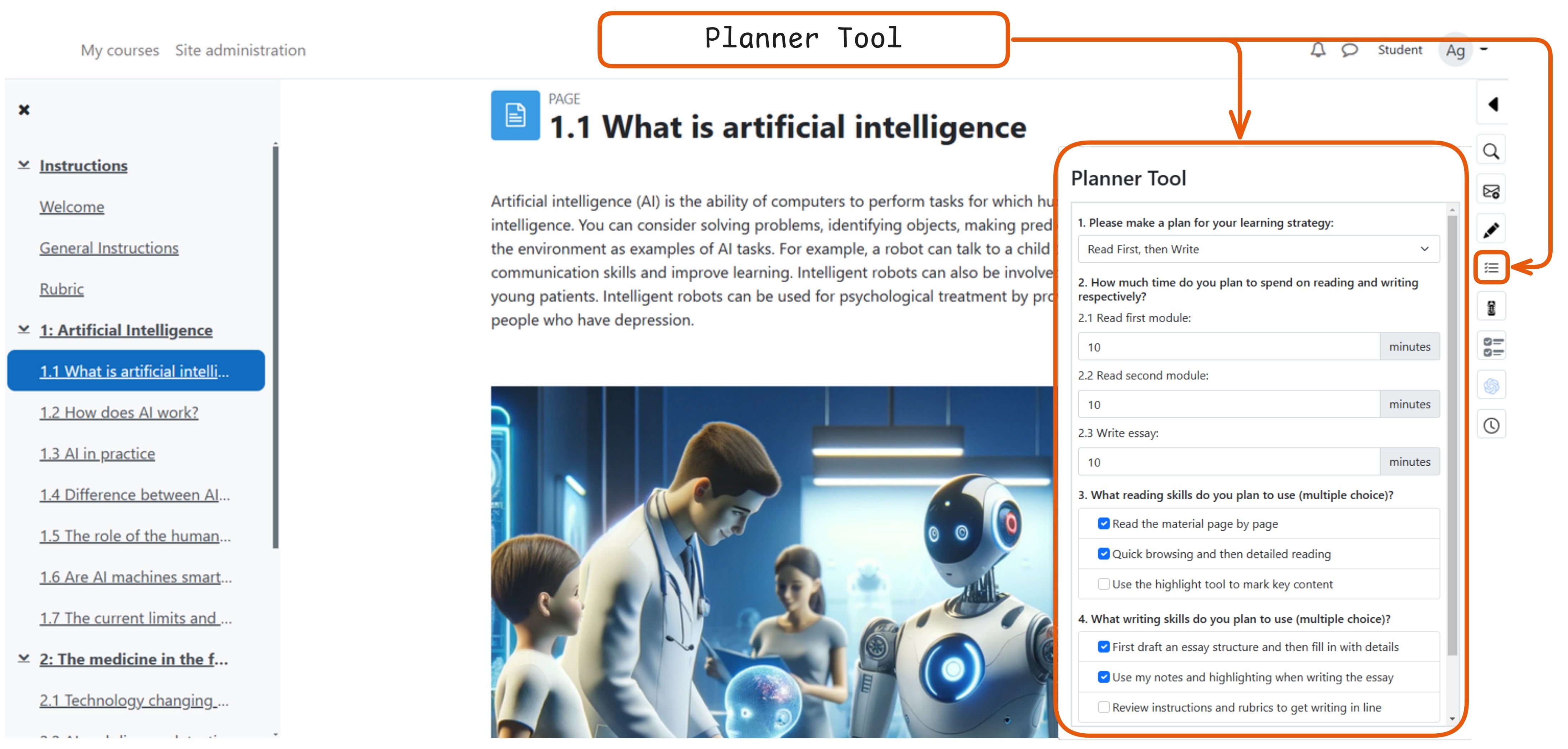}
\caption{Planner tool}
\label{fig:planner_tool}
\end{figure}

In the previous version of the \flora Engine, a planner tool was incorporated to capture students' planning actions during the learning process. Based on the existing studies, we found that few students chose to utilise the planner in their tasks. Feedback indicated that the planner was not sufficiently useful, and when tasks were short, creating a plan without any guidance was challenging and time-consuming. Drawing insights from a prior study~\citep{srivastava2022effects}, we observed that students' learning strategies can be primarily categorised into three main types: "Read first, then write," "Read and write simultaneously," and "Write intensively while reading selectively." Inspired by these analytical results, we upgraded the planner tool (as shown in Figure~\ref{fig:planner_tool}) by providing a step-by-step guide. This enhancement aims to make the planning process more intuitive and efficient for students. With the revised planner tool, students begin by selecting their main learning strategy, such as "Read first, then write." They can then allocate time to several predefined tasks and choose their preferred reading and writing strategies. By following this guided approach, students can complete their planning process within 2–3 minutes. This streamlined method not only makes the planner more accessible and user-friendly but also encourages students to engage in effective planning, even for shorter tasks.

\begin{figure}[ht]
\centering
\includegraphics[width=0.85\linewidth]{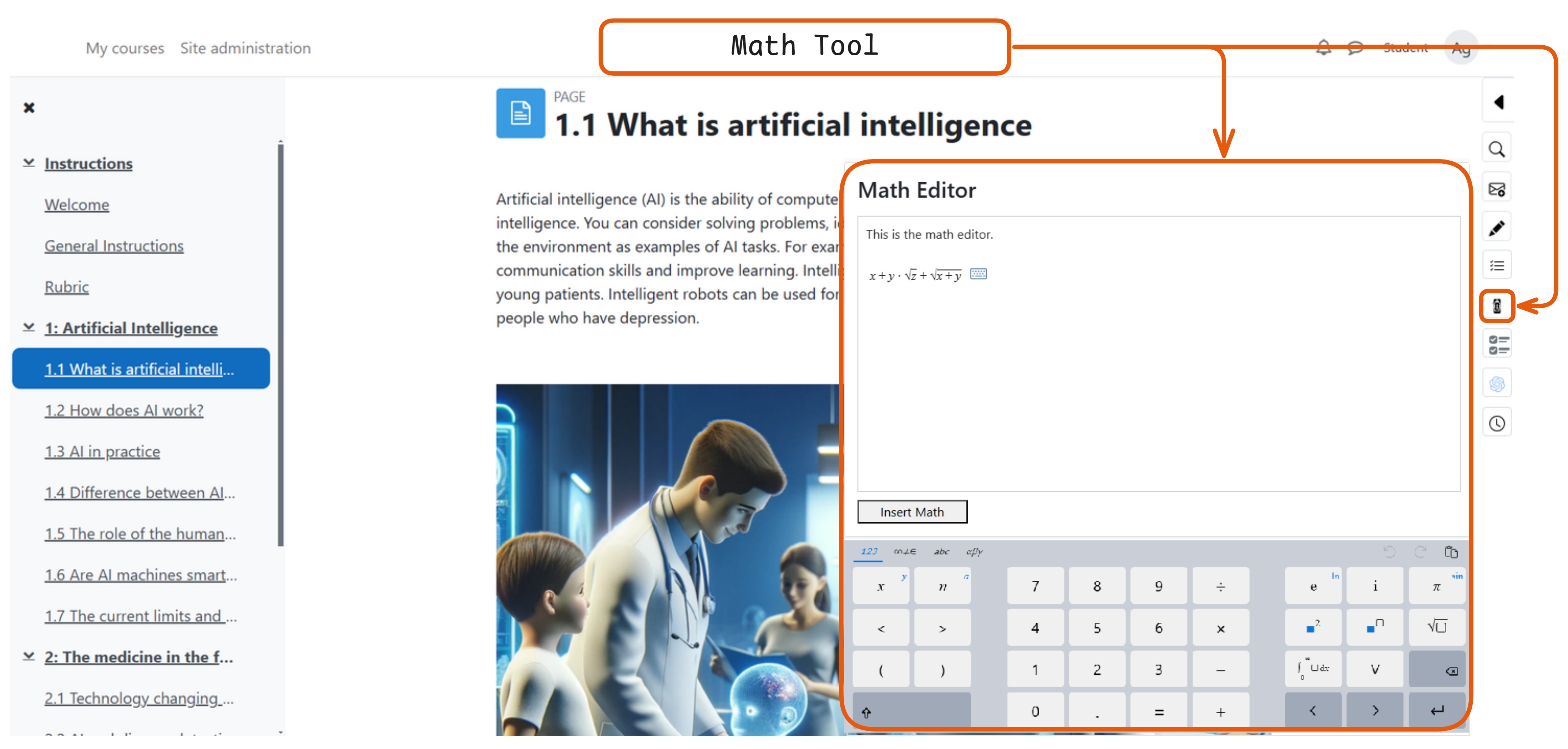}
\caption{Math tool}
\label{fig:math_tool}
\end{figure}

In addition to the planner tool, \flora now has a tool for mathematics (displayed in Figure~\ref{fig:math_tool}) to further expand the application scenarios of the \flora Engine beyond the reading and writing tasks. This feature allows learners to edit mathematical expressions directly within the platform, facilitating the integration of complex mathematical content into learning activities. By supporting the creation and manipulation of mathematical notation, the maths tool enhances the \flora Engine's functionality for STEM disciplines, such as mathematics and physics. This advancement broadens the platform's utility, enabling it to cater to a wider range of academic subjects and learning needs.

In addition to the aforementioned tools, the \flora Engine also offers several complementary tools designed to support diverse learning contexts. For instance, the integrated Dictionary tool leverages the Google Translation Service to facilitate multilingual translation, enhancing accessibility for users from different linguistic backgrounds. The text submission tool allow leaners to submit text to server. The Instant Chat tool allows learners to communicate directly with teachers, fostering timely interaction and support. Furthermore, the Questionnaire tool enables the administration of surveys or questionnaires during the learning process, thereby supporting data collection and formative assessment. These supplementary tools are frequently employed as complementary resources in specialised educational settings or are utilised as control group instruments in experimental studies. Due to their straightforward interfaces and minimalist, single-window designs, these tools are not depicted here in order to conserve space.

\subsection{GenAI-based Scaffolding Triggering Logic}\label{appendix_GPT_scaffold_trigger_logic}
This table describes the triggering logic of GenAI-based scaffolding tool.

\begin{table*}[p]          
  \renewcommand{\arraystretch}{1.5}
  \centering
    \caption{Learners' learning conditions ~\citep{li2025turning}}
  \label{tab:conditions}
  \begin{tabular}{|p{2.75cm}|p{6.4cm}|p{5.8cm}|}
    \hline
    \textbf{Condition Type} & \textbf{Learning Condition} &
    \textbf{How the Conditions were Captured} \\
    \hline
    \multirow{6}{*}{\textbf{Dynamic conditions}} &
      Strategic plan-making &
      action detection: SAVE\_PLANNER \\ \cline{2-3}
    & Awareness of the time constraint &
      action detection: TIMER \\ \cline{2-3}
    & Awareness of the available instrumentation tools &
      action detection: TRY\_OUT\_TOOLS \\ \cline{2-3}
    & Awareness of the available reading material &
      action detection: PAGE\_NAVIGATION \\ \cline{2-3}
    & Awareness of the task requirement &
      action detection: TASK\_REQUIREMENT \\ \cline{2-3}
    & Awareness of the available marking rubric &
      action detection: RUBRIC \\
    \hline
    \multirow{2}{*}{\textbf{Static conditions}} &
      Level of knowledge of learning strategies &
      ISDIMU~\citep{bannert2021isdimu} questionnaire score \\ \cline{2-3}
    & Level of prior knowledge &
      Pre-test score \\
    \hline
  \end{tabular}

\end{table*}

\bio{}
\endbio

\bio{}
\endbio

\end{document}